# Whispering gallery mode microresonator for nonlinear photonics

*Yihang Li, Xuefeng Jiang, Guangming Zhao, and Lan Yang*[*]

*Department of Electrical and Systems Engineering, Washington University, St. Louis, MO 63130, USA*
∗*Corresponding author: e-mail:* [yang@seas.wustl.edu](yang@seas.wustl.edu)

**Abstract:** Whispering gallery mode (WGM) microresonators, benefitting from the ultrahigh quality ($Q$) factors and small mode volumes, could considerably enhance the light-matter interaction, making it an ideal platform for studying a broad range of nonlinear optical effects. In this review, the progress of optical nonlinear effects in WGM microresonators is comprehensively summarized. First, several basic nonlinear effects in WGM microresonator are reviewed, including not only Pockels effect and Kerr effect, but also harmonic generations, four-wave mixing and stimulated optical scattering effects. Apart from that, nonlinearity induced by thermal effect and in PT-symmetric systems are also discussed. Furthermore, multistep nonlinear optical effects by cascading several nonlinear effects are reviewed, including frequency comb generations. Several selected applications of optical nonlinearity in WGM resonators are finally introduced, such as narrow-linewidth microlasers, nonlinearity induced non-reciprocity and frequency combs.

# 1. Introduction

Optical whispering gallery mode (WGM) microresonators have been actively studied since the 1980s. Starting from the droplet microresonators [1–3], WGMs were observed in not only fused silica microresonators, including microspheres [4,5], microtoroids [6–9], and microbottles/bubbles [10,11], and microdisk resonators [12–14]; but also polished crystalline microresonators [15,16], and integrated micro-ring resonators [17–19], which have attracted increasing attention in a variety of photonics applications, such as microlaser [7,20–29], optomechanics [30–34], bio/chemical sensing [14,35–49], nonlinear optics [50–52], chaotic photonics [53–67], non-Hermitian physics [68–75], *etc*. Utilizing methods like surface tension shaping and other optimized fabrication techniques, the surface scattering loss of a WGM could be significantly restrained, leading to ultra-long photon confining time, *i.e.*, ultra-high quality factor ($Q$). Meanwhile, WGM microresonators could achieve such confinement within a small mode volume ($V$). Therefore, the optical field of a WGM could be extremely intense, allowing observations and manipulations of nonlinear optical effects with a continuous-wave pumping. Compared with other types of resonators, such as photonic crystal cavities, WGM resonators support high $Q$ modes within a broad band of frequency, which is essential for nonlinear effects involving optical frequency conversions.

Optical nonlinearities in WGM microresonators have been found fundamentally interesting and promising in the development of novel light source, biomedical sensor, optical clocks, *etc*. Lots of nonlinear materials and structures were reported to improve the performance, and meanwhile, novel mechanisms like PT-symmetry breaking were introduced to interact with nonlinear phenomena in WGM microresonators. In this paper, we review the progress of nonlinear optics in WGM microresonators over the past three decades. This review is organized as follows. In Chapter 2, we provide a brief introduction on nonlinear effects in the WGM resonator, including some fundamental parameters as well as the methods to achieve the phase matching. In Chapter. 3, we review the progress of different types of basic nonlinear effects in WGM microresonators. In Chapter. 4, we discuss the recent advance of multistep nonlinear effects in WGM resonators, covering frequency comb generation and Raman-assisted processes. Chapter 5 summarizes the potential photonic applications. Finally, a summary and an outlook for future researches in this growing field are presented.

# 2. Nonlinear effects in WGM resonator

Under the spherical coordinate framework, a WGM could be characterized with three mode numbers ($n$, $m$, $l$). In spheres and spheroids resonators, an analytical description of the transverse component of WGM ($E_\phi$ or $H_\phi$) could be derived as [76]:

$$\psi_{nlm} = A\psi_r(r)\psi_\theta(\theta)\psi_\phi(\phi) \tag{1}$$

where $A$ is the mode amplitude and $\psi_r(r) = j_l(kr)$ (inside the resonator), $\psi_\theta(\theta) = P_l^m(\cos\theta)$ and $\psi_\phi(\phi) = \exp(\pm im\phi)$. Here, $k$ is the effective wave number of the WGM, determined by ($n$, $m$, $l$);

$j_l(kr)$ is the spherical Bessel function; and $P_l^m(\cos\theta)$ is the Legendre polynomial. Another set of mode numbers (m, p, q) could be introduced for a more straightforward description of the mode distribution. $p = l - |m|$, is the number of nodes in the polar direction. $q = n$, is the number of nodes in the radial direction. The resonant frequency of a WGM is explicitly related to the azimuthal mode number m:

$$\omega_m = \frac{mc}{R n_{eff}} \quad (2)$$

where $n_{eff}$ is the effective refractive index of the mode. In principle, $n_{eff}$ varies with the resonant frequency, leading to a deviation from an equidistant mode spectrum. Such a dispersive feature could hinder the effective production of a number of nonlinear effects. The component of the dispersion in WGM and the method for dispersion engineering will be discussed in Sec. 3.4.2.

Nonlinear optical effects are the optical effects where the optical properties of a system are modified by the presence of optical fields. Originating from the material response to the electromagnetic field, the nonlinear interaction could be characterized by the polarization $\tilde{P}(t)$ [77].

$$\tilde{P}(t) = \epsilon_0 \left[ \chi^{(1)} \tilde{E}(t) + \chi^{(2)} \tilde{E}^2(t) + \chi^{(3)} \tilde{E}^3(t) + \cdots \right] \quad (3)$$

In Eq. 3, the first term is the linear response of the material, and $\chi^{(1)}$ is the linear susceptibility in the linear optics. The latter terms refer to the nonlinear response, with the superscript representing the exponential of the electric fields. The origin of nonlinear susceptibility is comprehensively introduced in [77]. Here, we only treat those as intrinsic optical properties of materials. It should be noted that for the crystalline materials, susceptibilities usually have a tensor nature, leading to an anisotropic nonlinear response. For a $\tilde{E}(t)$ oscillating at optical frequencies, the nonlinear terms could cross-link different frequency components, causing modulations, conversions, and oscillations among the spectrum.

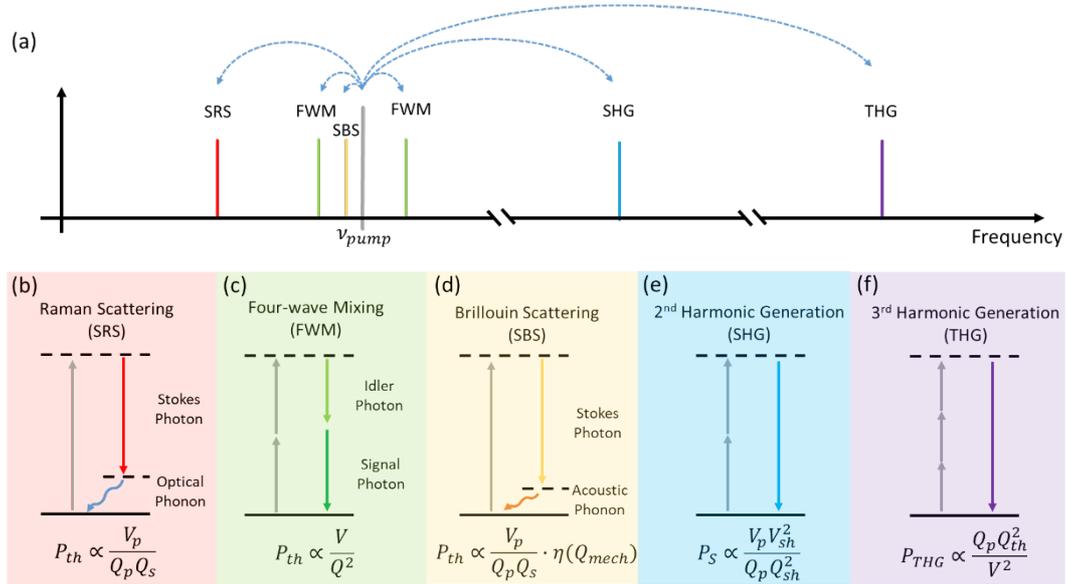

Fig. 1. Schematics of nonlinear wavelength conversions and corresponding state paragraphs. Underlying expressions are the relation between the nonlinear performances vs. resonator parameters. $P_{th}$ is threshold pump power; $P_S$ is the pump power at the conversion saturation, and $P_{THG}$ is the output power of third harmonic generation.

$Q$ and $V$ are quality factors and mode volumes, with subscripts representing different modes. For four wave mixing (FWM), because pump, idler and signal modes have rather close frequencies, Q and V are considered to be shared by all modes.

Frequency, or wavelength, conversions are the most significant effect in nonlinear optics. In Fig. 1, all of the building blocks of wavelength conversions are summarized and shown by energy level diagrams. Those conversions could be classified by whether to be parametric. If there is no energy or momentum exchange between light fields and the material, *i.e.*, the material stays in the original state after the conversion, the process is known as a parametric process, like those shown in Figs. 1(c), 1(e), and 1(f). On the other hand, stimulated optical scattering processes, including both stimulated Raman scattering (SRS) (Fig. 1(b)) and stimulated Brillouin scattering (SBS) (Fig. 1(f)), are not parametric. During the optical scattering processes, with the generation of phonons are also excited in the material, promoting the original state to a new virtual state. The curvy arrows refer to the phonon damping inside the materials, which is a follow-up process independent to the corresponding nonlinear optical effects.

If counting both photons and phonons participating in the processes, energy is conserved during the processes. On the other hand, the conservation of momentum is also critical for nonlinear optical effects. These two requirements refer to the full conservation in the ($\omega$, $\boldsymbol{k}$) space. Only if the total momentum or the wavenumber $\boldsymbol{k}$ is conserved before and after the nonlinear interaction, or "phase-matched", the energy of conversion outcome could be efficiently accumulated. Otherwise, the conversion efficiency will be oscillatory. For simplicity, here we consider a three-photon interaction, where photon 3 is generated by photons 1 and 2 (*e.g.*, SHG in Fig. 1(e)).

In a WGM microresonator, the wavenumbers of optical modes are characterized by the effective indices $n_{eff}$, and are proportional to the azimuthal mode number $m$ according to Eq. 2. Therefore, the phase-matching condition could be translated into the conservation of mode number m in WGM resonators. For simplicity, we write down the analytical expression of a three-photon interaction considering the phase-matching condition, where photon 3 is generated by photons 1 and 2 (*e.g.*, SHG in Fig. 1(e)) [78].

$$\frac{dA_3}{dz} = \frac{i\chi_{eff}\omega_3}{k_3 c} A_1 A_2 e^{i\Delta m z/R} \qquad (4)$$

where $A_i$ is the light intensity; z is the space coordinate along the circulating rim of the WGM resonators; $\Delta m = m_1 + m_2 - m_3$ ; $\chi_{\text{eff}} = \omega_3 \chi^{(2)} \frac{\sigma_{123}}{\sqrt{\sigma_1 \sigma_2 \sigma_3}} \sqrt{\frac{1}{2\varepsilon_0 c^3 n_1 n_2 n_3}}$ ; $\sigma_i = \iint \psi_i^2 dS_{\text{transverse}}$ and $\sigma_{123} = \iint \psi_1 \psi_2 \psi_3 dS_{\text{transverse}}$.

According to Eq.4, an efficient nonlinear effect requires two criteria: i), critical phase-matching (CPM) condition, *i.e.*, $\Delta m = 0$; ii) a large mode overlap, achieved by maximizing $\sigma_{123}/\sqrt{\sigma_1 \sigma_2 \sigma_3}$. The main advantage of using resonant structure to observe nonlinear optical effects is the strongly built-up optical field in the cavity, *i.e.*, enhanced $A_1 A_2$. The optical field enhancement could be characterized by the build-up factor, which is proportional to $Q/V$ of the corresponding mode. In the nonlinear optical effects, optical fields at various frequencies are usually involved, resulting in different scales of resonant enhancement. In Fig. 1, different relations between the resonators' parameter and the theoretical performance of nonlinear optical effects (efficiency or threshold) are summarized.

All expressions in Fig. 1 are based on the assumption of CPM. In practice, CPM could be achieved by leveraging birefringence or high-order transverse mode, so that the effective indices of corresponding modes could be aligned. On the other hand, for WGM based three-wave-mixing, quasi-phase-matching (QPM) is also valid, in which the susceptibility is modulated periodically along the rim to compensate the phase mismatch. In Secs. 3.2 and 3.4, more detailed schemes to achieve phase-matching condition will be discussed.

For a hyper-parametric process, like four-wave mixing and frequency comb generation, the phase-matching condition turns into a requirement of local dispersion rather than the index alignment between distantly separated frequency bands. Under such a circumstance, the selection rule of *m* and the energy conservation infer the equidistance of modes among the local band. It can be characterized by the group velocity dispersion (GVD) or group velocity dispersion parameter (D). While the theory of dispersion is mainly inherited from the field of fiber optics, those parameters could be translated to fit in WGM case [68]:

$$GVD = \frac{d^2k}{d\omega^2} = -\frac{1}{4\pi^2 R} \cdot \frac{\Delta(\Delta\nu_m)}{(\Delta\nu_m)^3} \quad (5.1)$$

$$D = \frac{d^2k}{d\lambda d\omega} = \frac{\nu_m^2}{2\pi Rc} \cdot \frac{\Delta(\Delta\nu_m)}{(\Delta\nu_m)^3} \quad (5.2)$$

where $\Delta\nu_m$ is free spectral ratios (FSR) with FSR = $(\nu_{m+1} - \nu_{m-1})/2$, and $\Delta(\Delta\nu_m) = \nu_{m+1} - 2\nu_m + \nu_{m-1}$. Equidistance requires *GVD* or *D* equals to zero, and *GVD* > 0 (*D* < 0) and *GVD* < 0 (*D* > 0) refer to normal and anomalous dispersion, respectively. In Sec. 4. 1, detailed way to engineer dispersion will be introduced.

## 3. Basic nonlinear effects in WGM microresonators

### *3.1 Pockels effect*

Pockels effect, also known as the linear electro-optic (EO) effect, is one of the most basic and simplest nonlinear optical effects. As a second-order nonlinearity, it describes the refractive index changing that is linearly proportional to the externally applied electric field, denoted as [80]:

$$n(\boldsymbol{E}) \cong n - \frac{1}{2}n^3 r \boldsymbol{E} \quad (6)$$

The coefficient *r* is the Pockels coefficient, and the applied electric field could be either DC or oscillating at a radio frequency. In most crystalline nonlinear materials, *r* is a tensor.

Pockels effect is the dominant mechanisms for the EO phase modulation in optical communication. In a resonator, phase modulation will lead to a shift of resonance and achieve amplitude modulation in the output signal (Fig. 2(a)). It should be noted that the resonator not only translates the phase change into the amplitude change but also shrink the necessary interaction length to a micrometer level. The key performances to evaluate an EO modulator are energy consumption and modulation speed. The former one, with the dimension of energy per bit, could be calculated by $E_{\text{bit}} = 1/4\, CU^2$ [81], where $C$ is the net capacitance of the modulator, and $U$ is the amplitude of the applied voltage signal. Both the structure and the efficiency of a modulator should be optimized to reduce $C$ and $U$, respectively. However, as for optical nonlinearity research, a more commonly used figure of merit is the tunability, *i.e.*, the resonance shift divided by the applied voltage. It is worth noting that the high tunability itself could not guarantee a good efficiency, or small $U$, because an effective modulation requires a frequency/wavelength tuning of roughly a full width half maximum (FWHM) of the resonance. Thus, with the same tunability, a modulator could benefit from a high $Q$ resonator. On the other hand, a high $Q$-factor means a long field build-up time, which sets the upper limit for modulation speed. Similar with the power consumption, the modulation speed is influenced by two factors: resistance-capacitance (RC) time constant of the structure and the build-up time. As a reference, an ultra-high $Q$ factor of $10^6$ at 1550 nm refers to a photon lifetime about 0.8 ns, limiting the modulation speed under 0.2 GHz. Normally, to achieve a high-speed modulator with a speed exceeding Gb/s, the $Q$ factor of WGM should not exceed the $10^5$ level.

As the Pockels effect only involves the nonlinear interaction between light fields and low-frequency electric fields, there is no requirement for the phase-match condition. The tricky part to build EO modulator is the choice of materials. Common optical materials like silica, silicon, silicon nitride have zero second-order nonlinearity. Even though alternative mechanisms like free-carrier plasma dispersion was introduced to electrically tune silicon's refractive index [82,83], limitations like undesired absorption and the trade-off between tunability and speed still leave the field open for other materials. In the past decade, the research of EO modulator in microresonators mainly focused on how to fabricate high-quality optical devices out of second-order nonlinear materials and how to compact the optical devices with electronic controlling. In the following paragraphs, some representative works will be reviewed based on materials.

Lithium niobate (LiNbO3, LN), is the most commonly used nonlinear material for EO modulation in fiber optics [84]. It possesses not only a large Pockels coefficient ($r_{33}$=31pm/V, $r_{13}$=8pm/V), but also a wide transparency range (0.4 μm -5 μm) [85]. In Ref. [86], Guarino *et al.* demonstrated the first microring resonator fabricated from a sub-micrometer LN thin film. The LN substrate was prepared based on the technique of crystal ion slicing and wafer bonding [87]. A $Q$ factor of 4×10³ in C-band and a tunability of 0.14 GHz/V were reported. In recent years, high-quality LN-Over-Insulator (LNOI) wafers with similar techniques have been commercialized, triggering lots of relative works on integrated LN photonics. LN resonators with $Q$ factors exceeding $10^6$ and tunability ~ GHz/V were reported with conventional semiconductor fabrication technologies [88–90]. On the other hand, resonators with hybrid materials were also developed for compatibility with silicon platforms [91–94]. As shown in Fig. 2(c), LN layer was bonded adjacent to resonators with evanescent field coupling.

EO polymer has also been widely studied for the Pockels effect. By introducing chromophores into a wide variety of polymers, high Pockels coefficients over 100 pm/V was reported [95], and many polymer-based EO modulators were reported with superior performance [95–100] (See also in Table 1). Meanwhile, hybrid schemes were also employed, including silicon-polymer ring [89,101,102], metal-polymer plasmonic ring [103] and titanium dioxide-polymer ring [104]. Most polymers were fabricated by the sol-gel technique, providing better compatibility with hybrid structures. For example, silicon-based slot structures could be involved to enhance the light field in EO polymer [101,102] (Fig. 1(e)).

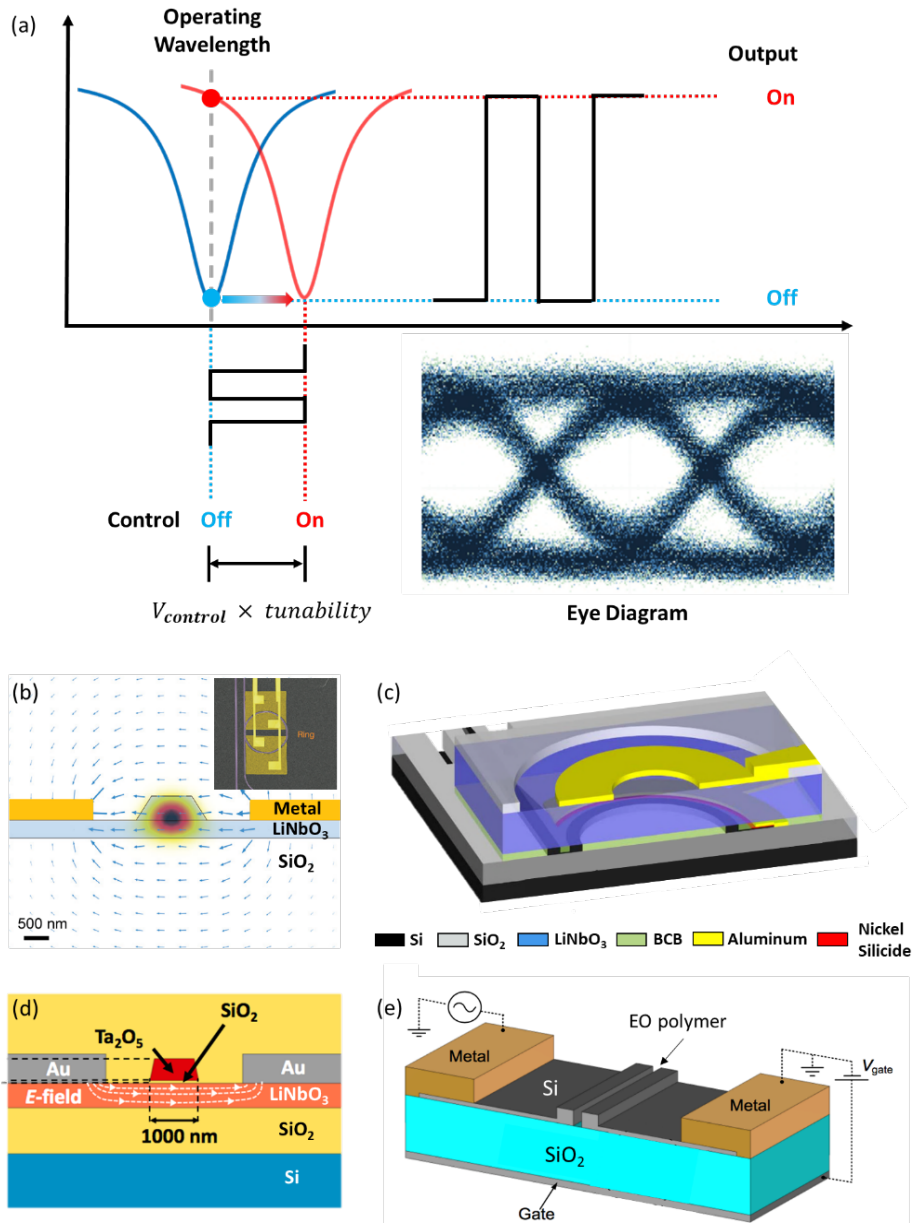

Fig. 2. Schematics of electro-optic (EO) modulation and representative configurations: (a) Working at the operation wavelength, control voltage shifts the resonant lineshape and modulates the output signal. Inset: A typical eye diagram of EO modulator based on lithium niobate (LN) ring resonator [89]; (b) Monolithic LN microring [89]; (c) Hybrid LN layer above the resonator [94]; (d) Hybrid LN layer below the resonator [93]. (e) Si-polymer hybrid configuration based on slot-waveguide [101].

In addition, various deposition techniques were employed to introduce nonlinear materials. In 2012, Chi Xiong *et al.* demonstrated AlN microring resonator for modulation out of AlN-on-silicon wafer through CMOS- compatible sputtering techniques [105]. In 2015, Bo *et al.* reported a LN-silica hybrid microdisk resonators, in which a 300 nm polycrystalline LN thin film was deposited on an inverted-wedge silica disk via excimer laser ablation technique [106].

Several aforementioned EO modulators' performances are summarized in the following table. Due to the enormous number of relative works with numerous different materials, a review covering most of those is not possible. Readers could refer to the cited paper for broader coverage.

Table 1. Parameters and performances of WGM-based EO modulator in different material platforms.

| Material | Structure | Q factor | Tunability (GHz/V) | Speed (GHz) | Power consumption (fj/bit) |
|---|---|---|---|---|---|
| Lithium niobate | Monolithic ring [89] | $5 \times 10^4$ | 0.85 | 44* | 240 |
| | Si-LN hybrid ring [94] | $1.4 \times 10^4$ | 0.42 | 9 | $4.4 \times 10^3$ |
| EO Polymer | Monolithic ring [96] | $6.2 \times 10^4$ | 0.82 | 2 | / |
| | Si-Polymer slot ring [102] | $5 \times 10^3$ | 1.63 | 1 | / |
| | Hybrid Polymer plasmonic ring [103] | $10^2$-$10^3$ | 3.6V for 3dB shift* | >100 | 5 |
| | TiO$_2$-Polymer hybrid ring [104] | ~$5 \times 10^3$ | 2.6 | 0.02 | / |
| III-V | Polycrystalline AlN ring [105] | $8 \times 10^4$ | 0.07 | 2.3 | 10 |

*Measured in another racetrack resonator with Q ~ 8000

Pockels effect could also be used as a detection mechanism of weak microwave signals. The index of the nonlinear resonator material could respond to the incoming microwave fields, which could be visualized by the probe light circulating in the resonator [107–114]. In this way, information in the microwave signal is up-converted to optical carriers for low-loss transmission, which is critical for quantum information networks with microwave qubits. In a WGM resonator, the ideal case would be the blue sideband of the microwave-modulated pump light resonant with an adjacent WGM but not for the red sideband. Up to now, three methods were reported to achieve single-band conversion. In

Ref. [111], the pump frequency was detuned from the WGM to introduce the frequency mismatch for down-conversion. In Ref. [110], asymmetric FSR was applied to suppress the down-conversion. In Ref. [111], mode splitting by coupled resonators was employed to address only the up-conversion.

*3.2 Three-wave mixing*

The second order parametric wavelength conversion refers to three-wave mixing, *i.e.*, light fields at three wavelengths (could be identical) $\omega_1$, $\omega_2$, $\omega_3$, where $\omega_3 = \omega_1 \pm \omega_2$, interact with each other with light generation. Different types of three-wave mixing were reported, namely second harmonic generation (SHG), where $\omega + \omega \rightarrow 2\omega$; sum/difference frequency generation (SFG/DFG), where $\omega_1 \pm \omega_2 \rightarrow \omega_3$; and optical parametric oscillations (OPO), where $\omega_3 \rightarrow \omega_1 \pm \omega_2$. For the two generated lights in OPO, the one with the higher frequency is called signal, the other idler. The generated signal/idler light could interact with the pump light through DFG and further generate idler/signal light, forming an oscillation with a threshold, beyond which the generation rates of both signal and idler could exceed the mode loss and build up. The frequencies of signal and idler could be tuned by controlling the phase-matching condition of signal and idler modes, thus allowing a broad dynamic range. OPO shows a great promise in the generation of coherent light [115], as well as squeezed light generation [116].

According to Eq. 4, three-wave mixing requires phase matching condition and non-zero mode overlap ($\sigma_{123} \neq 0$). In a WGM resonator. It could be expressed as [78]:

$$m_3 = m_1 + m_2 + M \qquad (7.1)$$

$$p_1 + p_2 + p_3 \text{ is even} \qquad (7.2)$$

$$p_1 + p_2 \geq p_3 \qquad (7.3)$$

The latter two rules refer to the non-vanishing mode overlap and could be satisfied by properly selecting transverse mode families. The first one, *i.e.*, phase matching condition, requires specific dispersion engineering. While $m_i$ is the azimuthal mode numbers of corresponding WGMs, M is the number of susceptibility periods along the circumference (if applicable). As mentioned in Chapter 2, phase matching could be achieved through critical phase matching (CPM), where M = 0, or quasi-phase matching (QPM), where M ≠ 0 for QPM.

For crystalline nonlinear materials, the birefringence nature of the crystals could be applied to achieve CPM. The non-diagonal components of susceptibility link light fields with different polarizations, and the difference of birefringent indices could compensate the material dispersion, matching the refractive indices. In the three-wave-mixing, type I phase-matching refers to that two lower-frequency modes have the same polarization and is perpendicular to that of the higher-frequency mode. While type II phase-matching refers to two fundamental modes having orthogonal polarization. In 2010, Furst *et al.* reported SHG [117] and OPO [118] in polished LN WGM resonator with type I phase matching, and the birefringent indices could be dynamically tuned by temperature and externally applied electric field. Apart from such dynamic tuning, the actual critically phase-matched modes could also be selected by geometric parameters of resonators or orientation of optic axis. In Ref. [119], Savchenkov *et al.* controlled the signal and idler frequencies by altering the thickness of the resonator. In Ref. [120–122], an oscillating refractive index was realized along the optical path, so that the type I phase matching could be achieved in some specific points cyclically. Another way to engineer the effective index of WGMs is by leveraging the geometric dispersion of high-order modes. Such a method was widely used in integrated microring resonators, reported with materials of GaN [123], AlN [124]. In LN, high-$Q$ WGM resonators have been made through conventional dry etching [90,125,126] or femtosecond laser micromachining [121,127,128], with experimental observation of SHG and SFG. However, accurate analyzing of critical phase matching in such a promising platform is still absent.

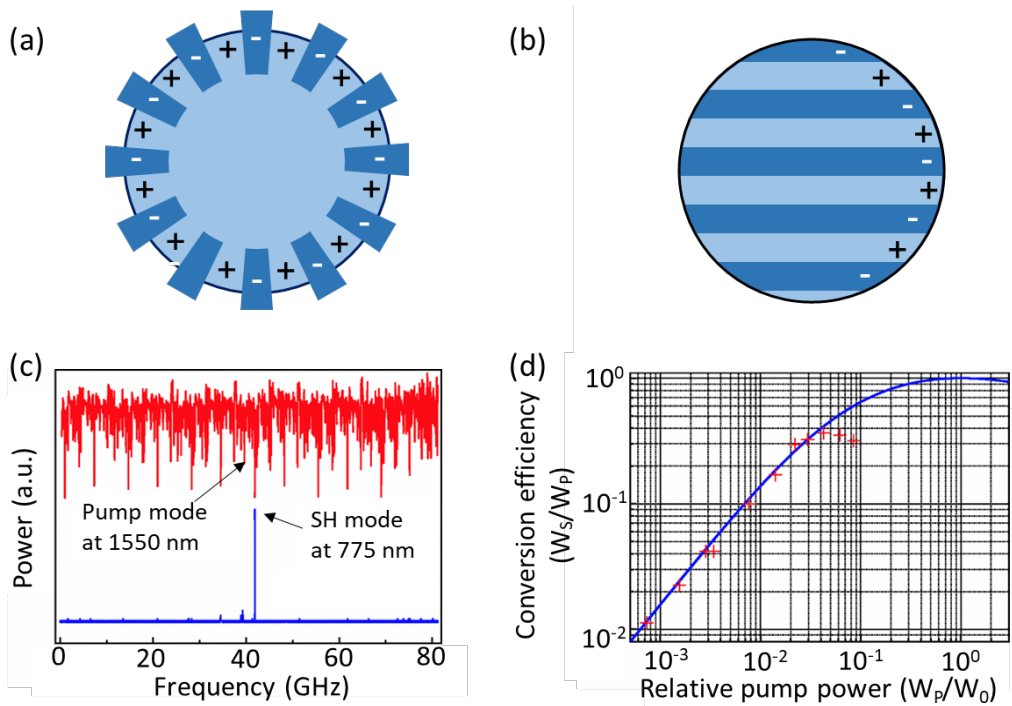

Fig. 3. Different poling patterns for quasi-phase matching (QPM) of LN disk, and second harmonic generation (SHG) performance (a) Radial pattern; (b) Parallel pattern. (c) Transmission spectrum near the pump mode at 1550 nm and generated SH mode at 775 nm. (d) SH conversion efficiency vs. pump power. A saturation phenomenon is observable [129].

QPM is based on the periodical modulation of the susceptibility, which could be realized by electric field poling or intrinsic crystal symmetry. Ferroelectric crystals, like LN, could be periodically poled to modify the nonlinear susceptibility, with techniques like calligraphic poling [130,131] and UV light assist [132,133]. A straightforward way of poling is shown in Fig 3(a). Such a pattern create a rigorous periodicity of susceptibility along the resonator rim. However, it also requires an accurate alignment between the center of the resonator and the poling pattern [133]. An alternative way is the parallel polling (Fig. 3(b)), which is easier to perform at a lower cost. Moreover, the varying period along the optical path could lead to a broad range of modulation Fourier component M, contributing to a broader dynamic range [129]. However, it is hard to apply material poling to devices smaller than millimeter size. For on-chip resonators, the symmetry of the semiconductor crystal is usually utilized for QPM. In $\bar{4}$ symmetry crystals like GaAs and ZnSe *etc.*, an intrinsic M of 2 is available in the plane perpendicular to the rotatory inversion axis. QPM from such symmetry was first studied theoretically [134] and experimentally realized by Kuo *et al.* in 2014 [135].

With a proper design for the phase matching condition, the performance of three-wave mixing could be improved significantly. In the following tables, the performance and the representative works of SHG and OPO will be summarized. For SHG, there are two types of conversion efficiencies which could be critical. At the low conversion efficiency limit (*i.e.*, low pump power), where the down conversion of second harmonic light is neglectable, SH Power is proportional to the square of the pump power. Therefore, a slope efficiency $\eta_s$, the increased efficiency per pump power, could be defined. When the phase-matching condition is well satisfied, the conversion efficiency could saturate at maximum value, denoted as $\eta_{\max}$, at the pump power of $P_S$ (Fig. 3(d)). $P_S$ is proportional to $\omega_p V_p^2 V_{sh}/Q_p^2 Q_{sh}$, and also determined by the nonlinear overlap between the pump and SH modes [47]. For OPO, the threshold power $P_{th}$ and the wavelength ranges of the signal and the idler are critical, both of which are determined by the phase-matching condition. Up to now, all experimental observations of WGM-resonator-based OPO were in LN resonators. Type-I CPM was applied when pumped in the visible range for both signal and idler in near-infrared [118,136,137], while quasi-phase matching fitted near-infrared pumping [78,138,139].

Table 2. Parameters and performance of second harmonic generation (SHG) in WGM resonators.

| Ref. | Material | D (μm) | Q factor | $\lambda_p$ (nm) | $\eta_s$ (%/mW) | $\eta_{max}$ (%) | $P_s$ (mW) | Phase matching |
|---|---|---|---|---|---|---|---|---|
| [129] | LN (B) | 3000 | 2×10$^8$ | 1319 | / | 2 | 30 | QPM(parallel) |
| [117] | LN(B) | 3800 | 3.4×10$^7$ | 1064 | / | 9 | 0.03 | Type I |
| [125] | LN(OC) | 28 | 1.0×10$^5$ | 1546 | 2×10$^{-4}$ | / | / | unclear |
| [121] | LN(OC) | 102 | 1.1×10$^5$ | 1540 | 1.1×10$^{-1}$ | / | / | CSPM |
| [120] | BBO(B) | 1820 | 7×10$^6$ | 974 | 4.6 | / | / | CSPM |
| [135] | GaAs(OC) | 5 | 3.3×10$^4$ | 1985 | 5×10$^{-3}$ | / | / | $\bar{4}$QPM |
| [123] | GaN(OC) | 80 | 1.0×10$^4$ | 1560 | 2×10$^{-5}$ | / | / | High-order modes |
| [124] | AlN(OC) | 60 | 2.3×10$^5$ | 1544 | 25 | / | / | High-order modes |

Table 3. Parameters and performance of optical parametric oscillation (OPO) in WGM resonators.

| Ref. | D (μm) | $\lambda_p$ (nm) | $\lambda_s$ (nm) | $\lambda_i$ (nm) | $P_{th}$ (μW) | Phase Matching |
|---|---|---|---|---|---|---|
| [119] | 870 | 1319 | ~1323 | ~ THz* | 2×10$^4$ | Type II |
| [118] | 1900 | 532 | 1010 | 1120 | 7-28 | Type I |
| [136] | 1900 | 532 | 1010-1060 | 1120-1060 | / | Type I |
| [137] | 1800 | 488 | 707-865 | 1575-1120 | 66-330 | Type I |
| [138] | 3100 | 1040 | ~2080 | ~2080 | 86-2200 | QPM(parallel) |
| [139] | 3080 | 1040 | 1780-2080 | 2500-2080 | 6000 | QPM(radial) |

## 3.3 Kerr effect

Kerr effect originally refers to the interaction between incident light fields and external low-frequency electric field (DC/RF). Given a light field of $E(t) = E_\omega \cos(\omega t)$ and a strong low frequency field electric ELF, the third order nonlinear polarization at ω reads $\tilde{P}^{(3)}(\omega) = 3\varepsilon_0 \chi^{(3)} |E_{LF}|^2 E_\omega \cos(\omega t)$. Compared with Pockel effect, the third-order susceptibility $\chi^{(3)}$ is usually much smaller than $\chi^{(2)}$ with the order of the characteristic atomic electric field strength (~$10^{11}$V/m) [77]. Hence Kerr effect may not be a good candidate for an efficient EO modulator. However, the term of non-oscillating $|E_{LF}|^2$ could be replaced the product of a pair of conjugate light field $E_{\omega'}^* E_{\omega'} = |E_{\omega'}|^2$. In this way, the controlling field is provided EM field from laser light and could be greatly enhanced in a high-$Q$ resonator. This phenomenon, known as optical Kerr effect, is an important tool for the all-optical signal processing.

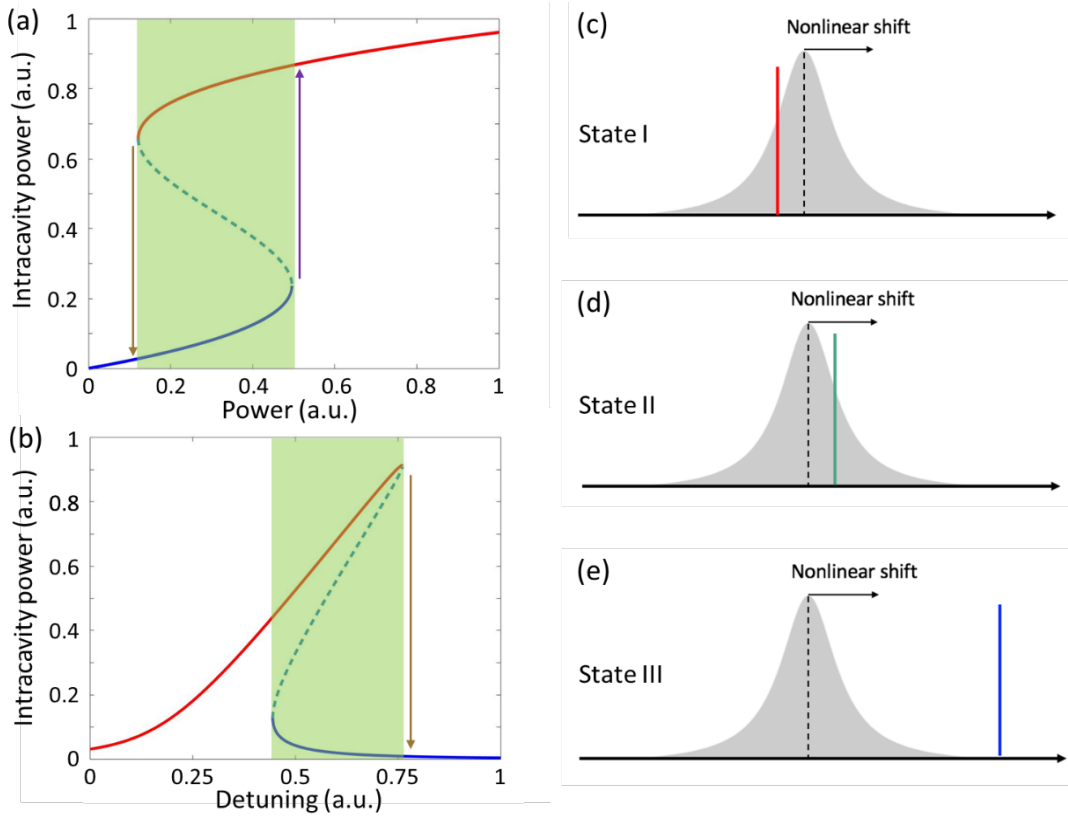

Fig. 4. Optical bistability of intracavity power with scanning pump power and pump detuning. Three-solution-regimes are shaded with green. (a) In power up-scanning, a switching on occurs at a boundary of three-solution-regime (purple arrow), while in power down-scanning, a switching off occurs at the other boundary. (b) Due to a strong enough pump, the original Lorentzian lineshape is deformed to be multi-valued. At the boundary of three-solution-regime, a sharp diving of intracavity power occurs, beyond which the resonance jump back to the original frequency/wavelength. (c) The relative position of the three solutions with respect to the transient WGM.

The direct results of optical Kerr effect are self-phase modulation (SPM) and cross-phase modulation (CPM), which refers to the optical Kerr effect on the pump field itself and another optical field, corresponding to $(\omega', k') = (\omega, k)$ and $(\omega', k') \neq (\omega, k)$, respectively. The optical Kerr effect could lead to a perturbation of refractive index, denoted as:

$$n = n_0 + n_2 I \quad (8)$$

where $n_0$ is the original refractive index, and $n_2$ is the nonlinear refractive index characterizing the strength of optical Kerr effect. For SPM and CPM, the value of $n_2$ differs by a factor of 2, as in $n_{2,SPM} = 3\chi^{(3)}/2n_0^2\varepsilon_0 c$; $n_{2,CPM} = 6\chi^{(3)}/2n_0^2\varepsilon_0 c$ [77]. In a WGM resonator, the intensity $I$ is largely enhanced, but the value is sensitive to the relative detuning between the pump and the resonant frequency, which is in turn tuned by the Kerr effect. The signature phenomenon of such a feedback loop is the optical bistability. As shown in Figs. 4(a) and 4(b), the optical bistability could be observed by scanning the pump power or pump detuning. Within the parameter space, three possible states could occur, as shown in Figs. 4c-4e. States I and III are stable, and state II is unstable. Sharp switching behavior could be observed at the transition points between three-solution-regime (shaded by green) and one-solution-regime [140,141]. Due to the stability of the three states, the hysteresis phenomenon will occur if reversing the scanning direction. It should be noted that those phenomena are not necessarily limited to SPM of the monochromatic input light. A dual-frequency input light of a pump and a probe is also available, where the pump shifts resonance for the probe via CPM [142–148].

In practice, however, the light intensity could also tune the resonance through the thermal-optic effect, which is usually much stronger than Kerr effect in WGM resonators. To observe the optical Kerr effect in WGM resonators, thermal effects must be suppressed by either environment cooling [140] or fast modulated input [141,142], since thermal response time is of the order of ms-μs, while Kerr effect is almost instantaneous. At room temperature, SPM and CPM have been observed in silica micro-toroid [142,146], silica bottle microresonators [141], polymer/protein coated microspheres [143,144], a:Si-H microresonators [145,148] and hybrid silicon-silica microsphere [147] and silicon carbide microdisks [149]. In addition, Kerr effect also showed a great promise in the realization of all-optical logic circuits [150–152].

### 3.4 Third-order parametric wavelength conversion

The third-order parametric wavelength conversion refers to the nonlinear interaction of four optical fields (could be degenerate) via $\chi^{(3)}$. Compared to three-wave mixing, the family of third-order parametric wavelength conversion has a much more diverse choice of the involved frequencies. In Sec. 3.4, two types of representative third-order parametric wavelength conversions are introduced, namely third harmonic generation (THG) and (degenerate) four-wave mixing (FWM). As shown in Fig. 1(c) and (f), THG and FWM refers to the wavelength conversion with three and two degenerate pump fields, respectively. They, together with the stimulated scattering processes, especially Raman scattering, could form a variety of cascade processes, which will be reviewed in Ch. 4.

### 3.4.1 Third-harmonic generation

In THG, three photons with identical frequency interact with each other and generate one photon with third harmonic frequency ($\omega + \omega + \omega \to 3\omega$). With a pump in the telecom band, the third harmonic will be visible light. Compared to SHG, THG has a much larger challenge for phase-matching engineering. Firstly, most materials have a rather strong material dispersion in the short-wave visible band. Secondly, a much larger wavelength difference between the pump and the third harmonic results in rather divergent mode profiles. Thus, a strong geometrical dispersion is also applied. Thirdly, in amorphous materials, neither birefringent CPM nor QPM is available, leaving the high-order mode CPM the only method for the phase matching engineering.

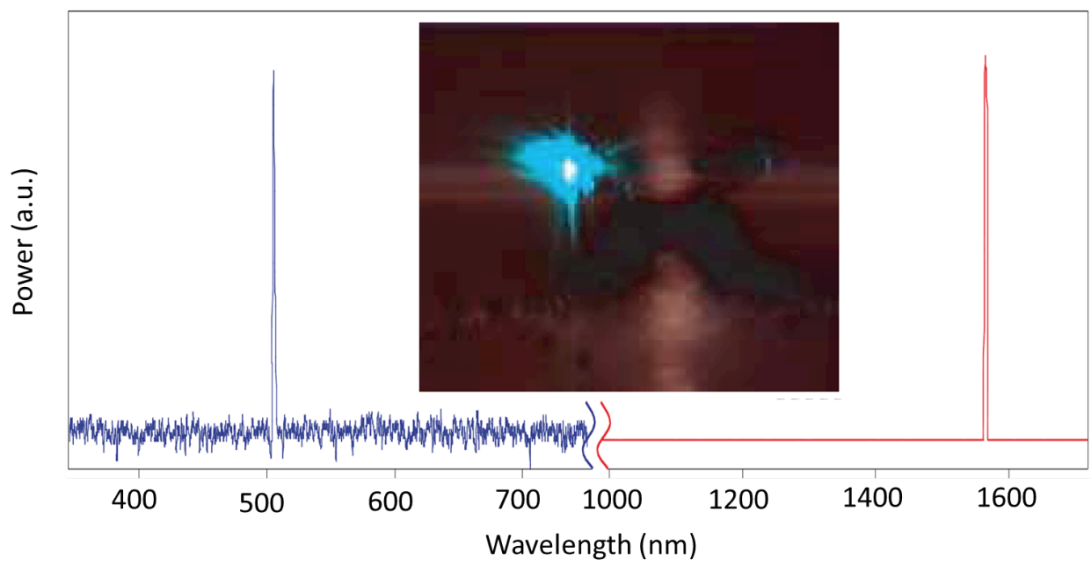

Figure 5. Third-harmonic generation in a silica microtoroid resonator [153].

The first observation of THG in WGM resonators could date back to 1980s. Chang *et al.* observed THG and Raman assisted third-order sum frequency generation (TSFG) in microdroplets made from $CCl_4$ and $D_2O$ [154–157]. In the last decades, THGs in solid-state resonators were reported in the silica microtoroid [153], microsphere [158], and silicon nitride microring [159], where high-order radial modes, polar modes, and waveguide-like modes were applied to achieve the phase-matching condition, respectively. It is worth noting that there is still plenty of room in the optimization of both conversion efficiency and collecting efficiency of THG in WGM resonators. The distinct difference between pump and harmonic modes limits the mode matching and efficient outward coupling. As a result, the collected signal was either from free-space collecting [154,155,157,160], or low-efficiency bus waveguide [153,158,159]. However, in 2017, Jiang *et al.* bypassed the phase mismatch of coupling by employing the momentum transformation via the chaotic modes in deformed silica microtoroids [53]. A 5000-fold improvement in device conversion efficiency was reported. In addition, THG could also be achieved by a cascade SHG and [161]. With QPM of LN applied, the external conversion efficiency could be improved.

*3.4.2 Four-wave mixing*

In this section, FWM specifically refers to the degenerate case, *i.e.*, the process where one idler and one signal photon are generated from two degenerate pump photons. FWM has a threshold effect in microresonators, above which the process is also known as hyper-parametric oscillations (HPO). FWM shows broad promise in applications, ranging from generation of non-classical state [162], all-optical frequency reference [163] (above threshold) to providing parametric gain for optical communication [164] (below threshold). Moreover, HPO could also serve as the building block of optical frequency comb generation, which is one of the research hotspots on nonlinear optics [165]. Compared to OPO, HPO does not rely on second-order nonlinearity. Thus, it is more generally exist in common optical materials like glass, silicon nitride, silicon etc. On the other hand, HPO involves WGMs that are spectrally close to each other (Fig. 6(a)) making the phase matching condition a local dispersion engineering problem rather than an index alignment at two distant frequency points.

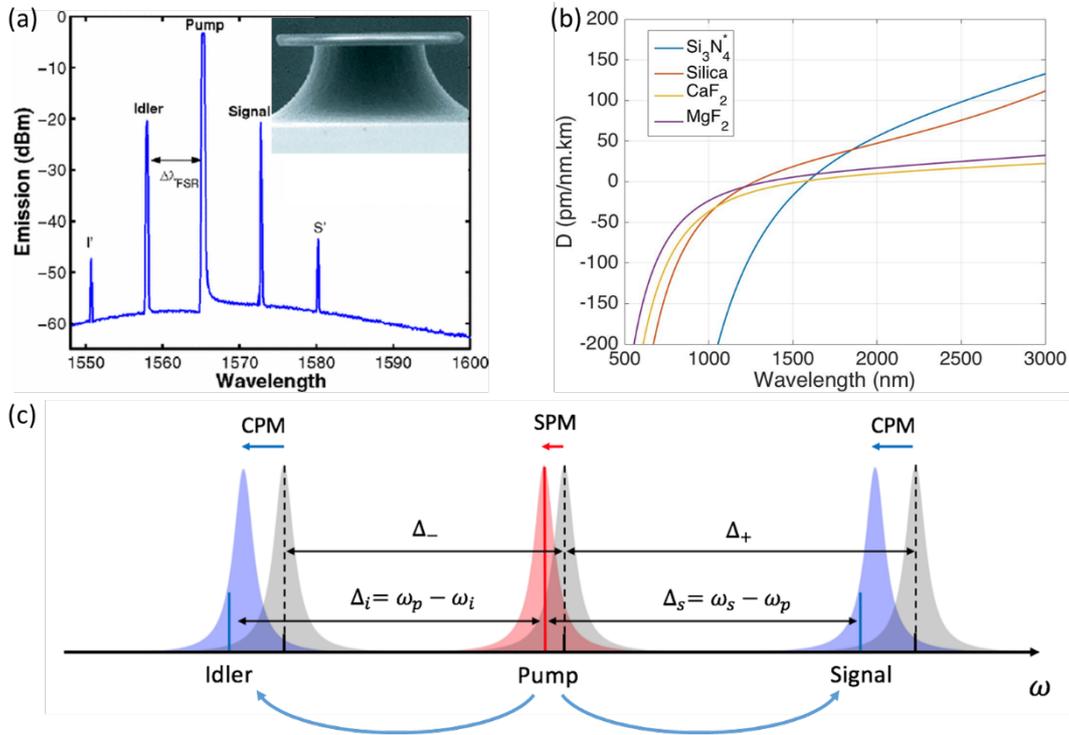

Figure 6. HPO in WGM resonators and dispersion engineering. (a) Spectrum of hyper parametric oscillation (HPO) in a microtoroid resonator [166]; (b) Material dispersion of typical nonlinear materials. (c) Schematics of mode pulling based on Kerr effect. $\Delta\_+$ and $\Delta\_-$ are the original mode spacing due to material and geometrical dispersion. $\Delta_i$ and $\Delta_s$ are the revised mode spacing due to the different pulling strength of self phase modulation (SPM) and cross phase modulation (CPM).

As stated in the Sec. 2, the phase-matching condition of FWM requires equidistant WGMs. This refers to a zero second-order dispersion (D = 0 or GVD = 0) near the center pump frequency. Within the relatively narrow band bounded by signal and idler frequencies, neither birefringence nor QPM is applicable for the dispersion engineering, leaving the geometric dispersion the major tool to compensate the material dispersion. The material dispersions of typical materials are shown in Fig. 6(b). Within the telecom band, fluorite glass has a low material dispersion, while fused silica and integrated silicon nitride has anomalous and normal material dispersion, respectively. The geometric dispersion of WGM resonators is determined by multiple factors varying with different resonator shapes. Unfortunately, there is not a versatile theory that could uniformly model the geometric dispersion in WGM resonators. Here, we will discuss several specific cases, but only with monolithic resonator configurations. In relative weakly confined structures like microspheres, microtoroids, and crystalline resonators, the geometrical dispersion usually provides a normal dispersion contribution that increases with the confinement, which could be tuned by changing the major radius and cross-section dimensions to fit different materials. At around 1550 nm, such a scheme has been employed to anomalous (or near zero) dispersion materials like silica [166,167], $CaF_2$ [163], $MgF_2$ [168]. Meanwhile, higher order polar or radial mode provides an alternative way to select the geometric contribution [168]. In structures like silica micro-bubble resonators with a dense high-order modes spectrum, such method is rather remarkable, where HPO at both telecom [169] and visible band [170] were observed. On the other hand, in strongly confined structures like micro-ring resonators, the small waveguide-like cross sections make their dispersion rather complex. With different thickness and width, the strip waveguide could have a hill shape in the total D spectrum [171], covering both normal and anomalous regions. Operating in different regions, FWM was achieved in silicon [172], gallium arsenide [173], silicon nitride [174,175], high index doped silica [176–178] and diamond [179].

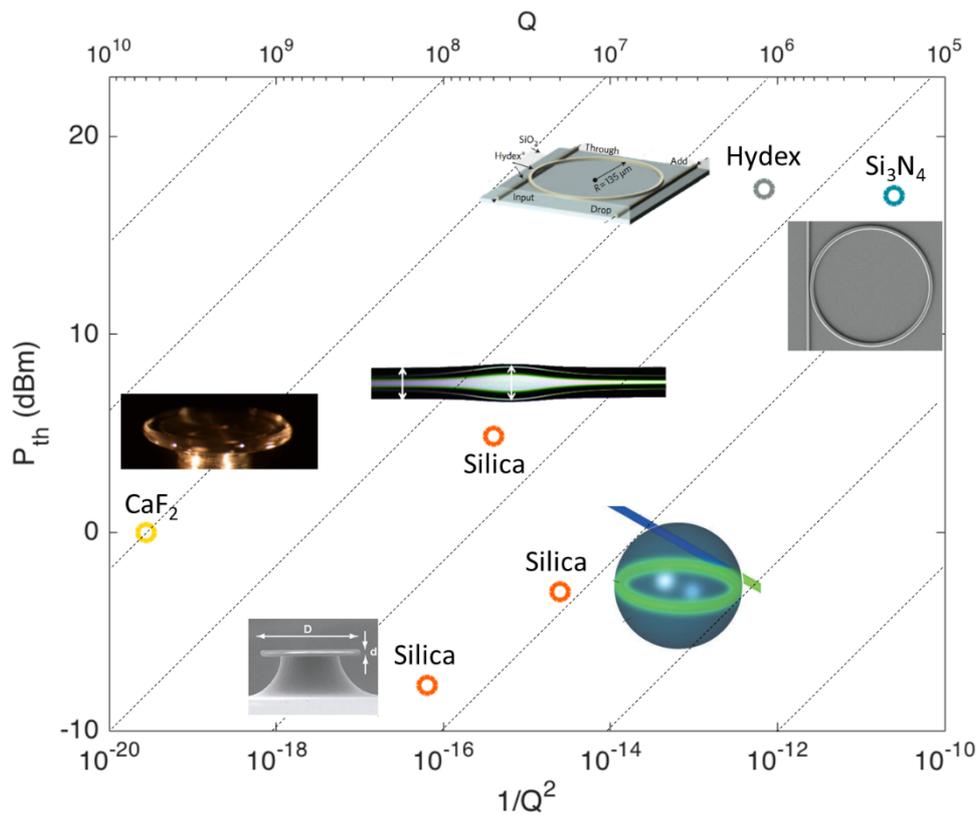

Fig. 7. Summary of HPO threshold in different WGM resonators.

However, dispersion engineering could only provide a rough tuning to the phase-matching condition, which is also sensitive to the fabrication process. On the other hand, Kerr effect could be employed for a fine and dynamic tuning of mode spacing. While both the pump resonance and idler/signal resonance are shifted by SPM and CPM respectively, their refractive index change differ by a factor of 2 ($n_2^{(cross)} = 2n_2^{(self)}$). As a result, idler/signal resonance are shift farther relative to the pump resonance, as shown in Fig. 6(c). Since most materials have a positive $n_2$, Kerr shift corresponds to a normal dispersion contribution. If the system has an originally anomalous total dispersion, *i.e.*, $\Delta_- < \Delta_+$, optical Kerr effect could "pull" the mode to the proper place, overlapping with an equidistant frequency triplet (lines in the figure). The maximum original mismatch $\Delta\omega = \Delta_+ - \Delta_-$ that Kerr mode pulling could fix is known as parametric gain bandwidth [166]:

$$\Omega = 4\frac{c}{n}\gamma P \qquad (9)$$

where $\gamma = \omega n_2/cA_{eff}$ is the nonlinearity parameter, P is the intracavity circulating power.

In experiments, when operating below the threshold, both a strong pump light and signal light are coupled into resonators, and the device works as a parametric wavelength converter [172,173,176]. Thanks to the strong enhancement within the resonator, an enhanced efficiency up to ~ -20 dB was observed with a rather low incident power of ~ 10mW [172,176]. The experimental demonstration of HPO was firstly reported in silica microtoroid [166], and then have been realized in CaF2 microdisk [163], silica microsphere [167], silicon nitride micro-ring [174], high-index doped silica micro-ring [177] and silica micro-bubble [169,170]. Most of the resonators had a rather high *Q* factor so that the single input at pump frequency could exceed the threshold. If neglecting the influence of non-ideal coupling, the threshold pump power reads [163]:

$$P_{th} \cong 0.77 \frac{\pi n^2 V}{n_2 \lambda Q^2} \qquad (10)$$

It should be noted that the expression has considered being under the phase-matching condition, without which HPO could be hindered by intrinsically phase-matched SRS (see Sec. 3.6.1) [166,167,180]. In Fig. 7, threshold powers of representative works are summarized. In such coordinates, points closer to lower right corner refers to high material nonlinearity, small mode volume or good phase-matching engineering, which could lower the threshold power.

### 3.5 Stimulated Scattering

In parametric conversions, nonlinear materials only provide parametric states to address the state transition without exchanging energy or momentum with photons. However, the fluctuation within the material at some intrinsic eigen-frequencies could, in fact, interact with the incident light. Such a process is known as scattering. In this section, we are going to review two types of wavelength conversion scattering processes in WGM resonators, namely Raman scattering and Brillouin scattering, which originate from the material fluctuations of different types. For Raman scattering, the fluctuation is the vibrational modes of the molecules constituting or optical phonons. For Brillouin scattering, light is scattered by the sound waves within the material, or acoustic phonons [77]. Scattering could create two sidebands of the pump, namely Stokes and anti-Stokes component, which losses and gains energy from fluctuations, respectively. In this review, we always focus on the Stokes component of the scattering process, whose state diagrams are shown in Fig. 1(b) and (d). The pump field promotes the atom state from the ground one to the excited one, and an existing Stokes photon could stimulate the transition between the excited states to the phonon state with the generation of another Stokes photon. Due to a relatively low lifetime of most of the involved phonons, a population inversion between the excited state and the phonon state could be achieved with a strong pump, forming Raman/Brillouin gain. In a WGM resonator, such population inversion is easier to achieve thanks to the light field enhancement, making it a good platform to observe and make use of SRS and SBS. Amplifiers and lasers could be demonstrated by SRS and SBS. Compared to conventional lasers, the absolute pump wavelength is not strictly required by the media. Hence, the laser wavelength could be selected more freely. What's more, SBS was found promising to achieve narrow-linewidth lasers, which will be reviewed in Sec. 5.1 in detail.

### 3.5.1 Stimulated Raman Scattering

In SRS, given an existing Stokes photon, a pump photon is converted into another Stokes, and promote the atom from the ground state a new one with an optical phonon excitation. Such a process contains both energy and momentum exchange between photons and phonons. Thus, the phase-matching condition should consider both photon modes and phonon modes. However, the photon momenta of both pump and Stokes light are much smaller than the boundary value of phonon's Brillouin zone. Thus, the process only involves the optical phonons with "near-zero momentum", where the dispersion of is rather flat. Therefore, while the frequency shift of light is determined by the phonon frequency of the flat area, the momentum conservation is easily satisfied due to the broadly available $k_O$ (subscript O for optical phonon). An alternative interpretation is based on the wave nature and a classical model of molecular oscillators, from which the Raman gain $g_R$ could be expressed [77]:

$$g_R = 3i \frac{\omega_R}{n_R c} \chi_R(\omega_R) |A_P|^2 \qquad (11)$$

where $\omega_R$, $n_R$ is the Raman frequency (Stokes) and the refractive index at the frequency; $\chi_R(\omega_R)$ is the effective susceptibility of Raman scattering; $A_P$ is the field amplitude of the pump light. It could be observed that such a coefficient depends only on the modulus of the pump field where all wavenumber terms $k_i$ (i: P, R, O) is absent, excluding the requirement of the phase matching condition.

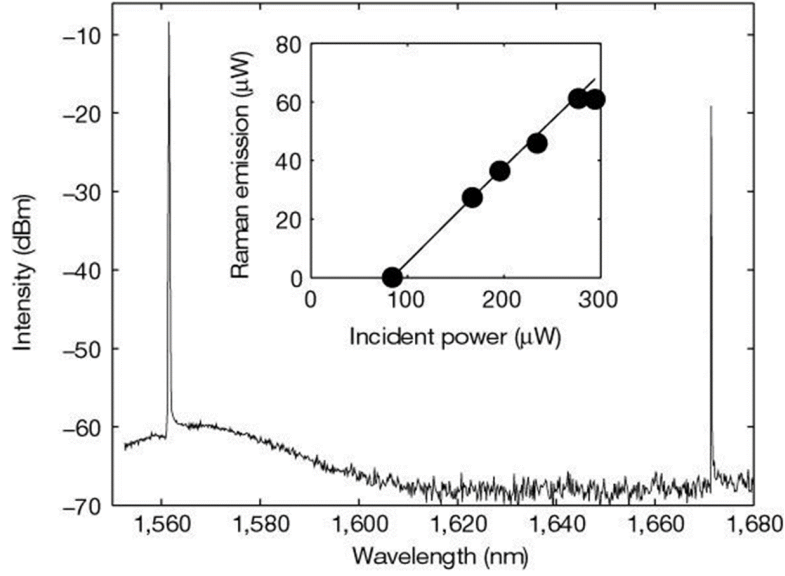

Fig. 8. Stimulated Raman Scattering (SRS) emission spectrum and threshold phenomenon [181].

In WGM resonators, the threshold of a Raman laser is [181–183]:

$$P_{th,R} = \xi \cdot \frac{\pi^2 n_P n_R V_P}{g_R \lambda_P \lambda_R Q_P Q_R} \quad (12)$$

where $Q_i$, $V_i$ refer to $Q$ factors and mode volumes, and $\xi$ is a factor accounts for the imperfection of coupling, mode overlap and backscattering. In the experiment, the early realization of Raman emission was observed in micro-droplets [2,184,185]. In 2002, Spillane *et at.* reported the first demonstration of Raman laser in a solid-state silica microsphere resonator [181]. When the pump power is far higher than the threshold, due to the rather wide Raman gain bandwidth of fused silica, multiple Raman peaks could be observed around the initial lasing wavelength. In Ref. [186], Kippenberg *et al.* reported the Raman lasing in fused silica microtoroid resonators with a smaller mode volume ($V_P$) and less degeneracy of polar modes. Thus, a single mode Raman laser is realized. Raman lasing was also reported in sol-gel based silica microtoroid by Yang *et al.* showing the versatility of the silica-based active device [26]. SRS could be further enhanced by involving another material with larger $\chi_R$. Hybrid structures including silica resonators coated with PDMS layer [187], Ti-doped sol-gel silica layer [188] and Zr-doped sol-gel silica layer [189] were reported, and the Raman lasing in chalcogenide glass $As_2S_3$ micro-sphere was also demonstrated [190,191]. Meanwhile, several novel structures for silica-based WGM Raman laser were employed. Ooka *et al.* reported Raman lasing in hollow bottle-shape WGM resonators with the high tunability [192]. Jiang *et al.* reported free-space coupled Raman laser in deformed micro-toroid [183,193]. In 2016, Zhao *et al.* realized Raman lasing in packaged microtoroids [194]. By freeing the devices from the translation stage for coupling fiber taper, such a platform is promising applications requiring portability and robustness. The record low threshold of Raman lasing was reported in Crystalline WGM resonators [182]. In a $CaF_2$ disk resonator with $Q$ factor as high as $5\times10^{10}$, a lasing threshold of 3 μW was achieved. Cascade Raman lasing, where the lower-order Raman light serve as the pump for the consecutive order, could also occur in WGM resonators. While in micro-droplets, a pulse laser is required to generate cascade Raman peaks [184], in solid-state resonators, sub-microwatt level pump power could multiple cascade peaks in silica [195,196] and CaF$_2$ resonators [182], respectively.

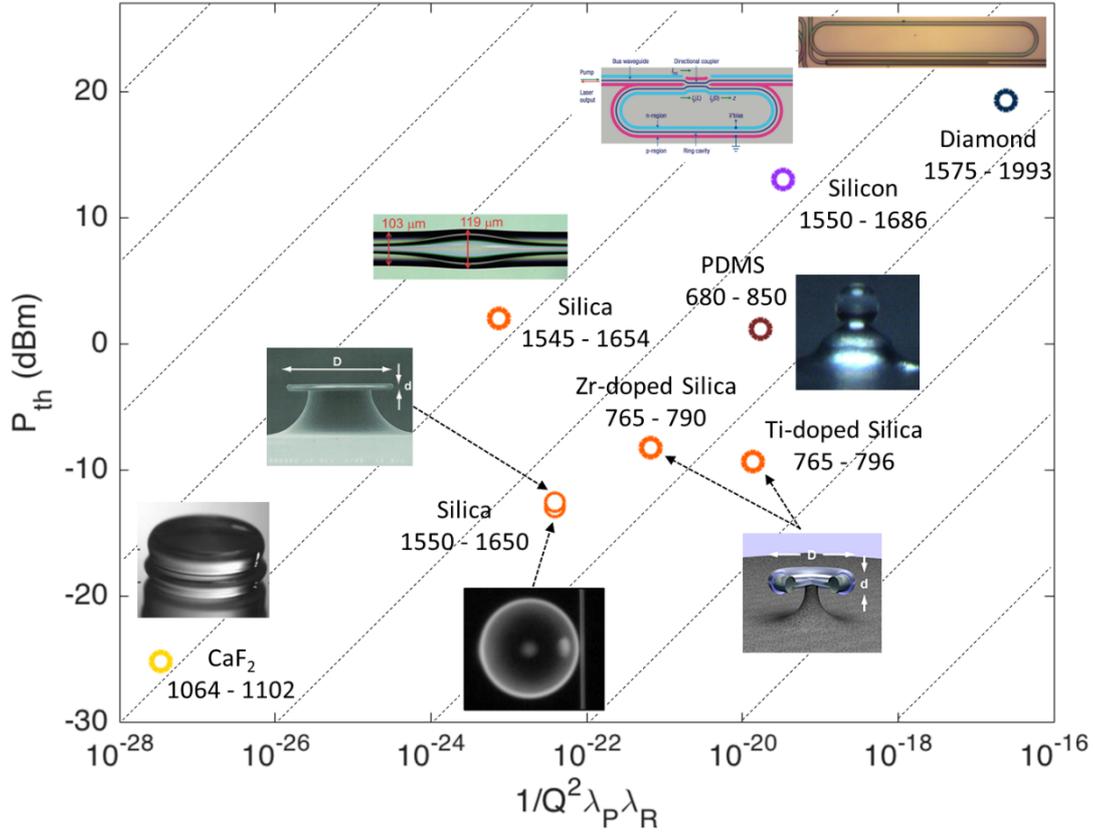

Fig. 9. Summary of SRS threshold in different WGM resonators (unit of pump and emission wavelength: nm)

Raman lasing in the integrated platform was also actively studied. Up to now, all the relative works are based on racetrack-shaped resonators instead of micro-ring resonators for a two-fold reason: 1. The straight coupling regime allows deterministic coupling engineering to achieve critical coupling at pump frequency and under coupling at lasing frequency (for lower threshold [186,197]); 2. The long straight regime could efficiently reduce the FSR in order to match the probably narrow Raman linewidth of some material [198]. In 2007, Rong *et al.* reported a continues-wave Raman laser based on a silicon racetrack resonator, with pumping at 1550 nm and lasing at 1686 nm [197]. The high nonlinear loss from the free carrier absorption was reduced by the p-i-n waveguide design, and tens of mW output were realized with a slope efficiency of 28%. In Ref. [199], a cascaded Raman laser (up to second order) was demonstrated in the same platform, where a critical coupling at pump frequency, near-zero coupling at first order Stokes frequency and weak coupling at second order Stokes frequency were achieved. Apart from silicon, diamond is also a promising material for integrated Raman laser due to its broad transparency window, large Raman shift (~40 THz) and Raman gain. In Ref. [198], Latawiec *et al.* reported a 1993 nm Raman laser with 1575 nm pumping on an on-chip diamond resonator. As shown in Fig. 9, we summarized the threshold powers of SRS lasers in different platforms. According to Eq. 12, $1/Q^2\lambda_P\lambda_R$ was selected as the x-axis, so that the points closer to the lower right corner refers to a higher Raman gain, a smaller mode volume or a better phase-matching.

SRS could also realize dopant-free active WGM resonators for different applications. In 2014, Ozdemir *et al.* [200] and Li *et al.* [201] [116] independently reported nano-sized particle sensing using active silica micro-toroids with Raman gain. When pumped below the threshold, SRS in WGM resonators could also work as amplifiers [202,203] and $Q$ factor enhanced resonators for sensing or optical control [200,204,205].

### 3.5.2 Stimulated Brillouin Scattering

Stimulated Brillouin Scattering (SBS) depicts the process of scattering between light and acoustic wave, or photon and acoustic phonon. During SBS process, the acoustic wave serves as a moving grating scattering the pump wave into a Brillouin wave, and the beating between the two optical waves reinforce the acoustic grating through the electrostriction effect. Different with optical phonon, acoustic phonons have nearly linear dispersion curve, *i.e.*, $\Omega_A = |\boldsymbol{q}_A|v$, where $\Omega_A$ and $\boldsymbol{q}_A$ are the frequency and wavenumber of acoustic phonon; and $v$ is the velocity of the acoustic wave in the media. Thus, it is critical for SBS to satisfy the phase-matching condition, considering both photon and phonon modes. With different acoustic wave travelling directions, two types of SBS are studied, namely backward scattering and forward scattering. As shown in Fig. 10(a), due to the energy and momentum conservations, different types have largely distinct Brillouin shift $\Omega_A$. Backward scattering has a much larger Brillouin shift (normally ~10 GHz) [206,207] than forward scattering (~10 MHz – 1000 MHz) [208]. However, neither of the shifts is big enough to match the FSR a submillimeter-sized WGM resonators. Thus, for an efficiency enhanced SBS in WGM resonators, it is critical to select two WGMs which are close enough to address the pump and Stokes frequencies. Moreover, WGM resonators may also support acoustic modes, which could further enhance SBS in resonators. This effect lead to an extra term in the threshold of SBS than that of SRS [206]:

$$P_{th,B} = \xi \cdot \frac{\pi^2 n_P n_B V_P}{g_B \lambda_P \lambda_B Q_P Q_B} \cdot \frac{1}{1+Q_A \lambda_A/2\pi R} \tag{12}$$

where $Q_i$ and $\lambda_i$ are the $Q$ factors and wavelengths, respectively. In the backward SBS, the GHz mechanical mode usually has a short lifetime, or a low $Q_A$, so that the second term converge to unity. In this way, SBS is rather a local phenomenon in the resonator, and the threshold power expression shares the form with that of SRS threshold power. However, the typical Brillouin gain $g_B$ is much larger (~100 time) than Raman gain $g_R$ [207]. Thus, as long as the phase-matching condition is satisfied, SBS has a lower threshold than SRS and FWM.

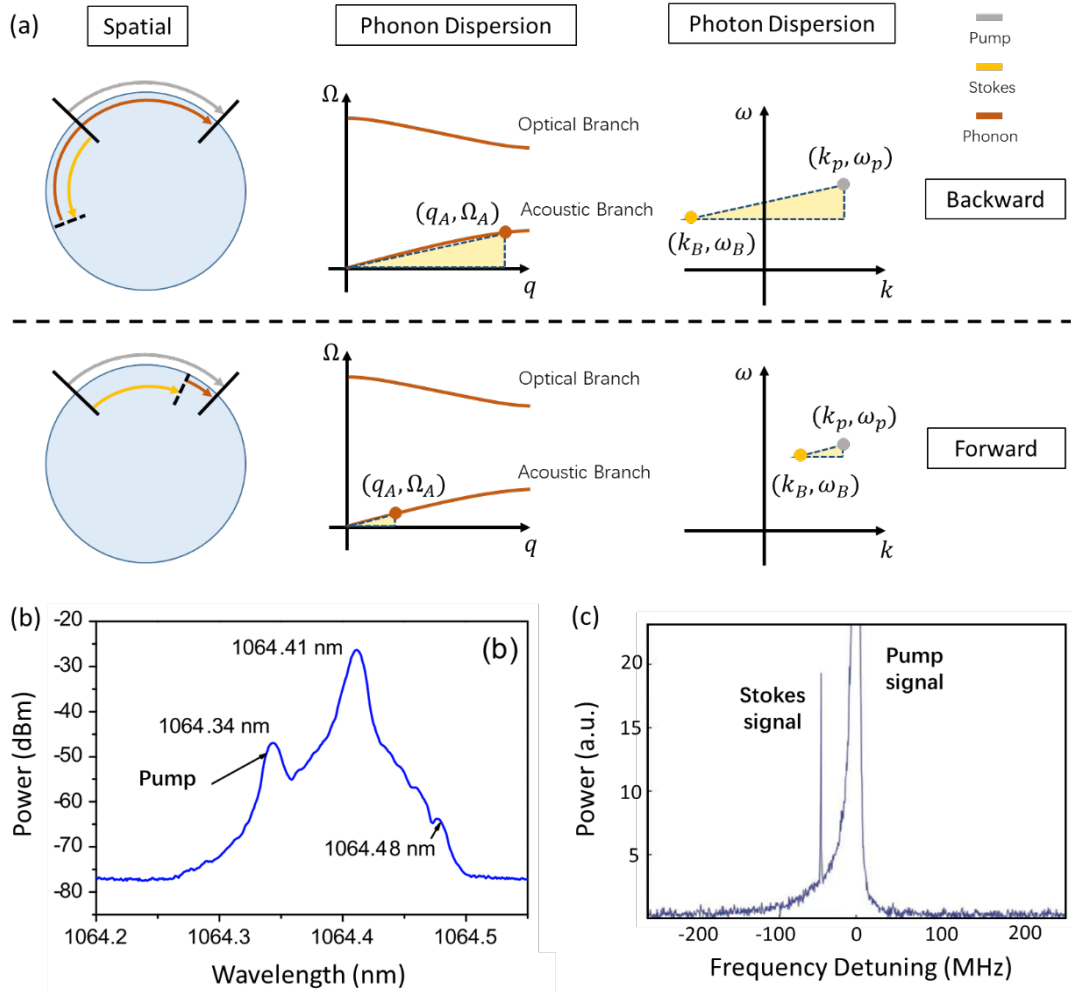

Fig. 10. Schematics and experimental spectra of backward and forward stimulated Brillouin scattering (SBS) in WGM resonators. (a) Yellow triangle regions refer to energy and momentum of corresponding phonons; (b) Reflective output spectrum of SBS in a CaF$_2$ resonator [207]; (c) Output spectrum of SBS in a silica microsphere [208].

Backward scattering was first observed in the experiment. According to the momentum diagram in Fig. 10(a), given the assumption of $v \ll c/n$ and the neglection of the refractive index dispersion, the Brillouin shift could be derived as [77]

$$\Omega_{A,backward} = \frac{2v}{c/n} \omega_P \qquad (13)$$

which agrees with the Doppler shift for a moving grating. Another statement associated with the assumption is $|k_A| \cong 2|k_P|$, or in azimuthal mode number: $m_A \cong 2m_P$. In Ref. [207], SBS lasing was first demonstrated in a CaF$_2$ disk resonator. The disk resonator had a major diameter of more than 5 mm, and have a dense high-order transverse mode ($p, q \neq 1$) to address the Brillouin shift of 17.7 GHz, and the ultra-high optical $Q$ factor ($4\times10^9$) gave rise to a record low SBS lasing threshold of 3 μW. SBS in amorphous silica micro-sphere was reported in the same year. In the sub-millimeter resonator, high-order transverse optical modes were applied to host the pump and Stokes signal, and a slope efficiency

of 90% was achieved [206]. In 2012, Lee *et al.* reported chemically etching ultra-high $Q$ wedge disk resonators, whose diameter could be accurately determined by lithographical techniques, so that the FSR could accurately match the SBS shift [13]. On the other hand, backward SBS lasing was reported in silica micro-bottle resonators [209,210], BaF2 crystalline resonators [211] and tellurite glass microspheres [212], where high-order transverse modes were applied to match the Brillouin shift. In all the platforms mentioned previously, cascade backward SBS could be observed. One unique feature for cascade backward SBS is the reversing of the output direction. Due to the backward emission, odd-order SBS occurs in the reflection direction, while even-order SBS occurs in the transmission direction.

Forward SBS has a much smaller Brillouin shift at MHz level, where the resonator usually support high-$Q_A$ acoustic mode. Thus, forward Brillouin scattering provides a promising platform to study travelling optomechanical behaviors at around Brillouin shift frequencies [213–215]. The Brillouin shift of forward SBS depend highly on the selection of the involved optical and acoustic modes [208]. In 2011, Bahl *et al.* demonstrated forward SBS in fused silica micro-sphere resonators [208], as shown in Fig. 10(c). Rayleigh-type surface-acoustic-wave was calculated to fit the phase-matching condition. By scanning the pump frequency, numerous mechanical modes with frequencies ranging from 49 MHz to 1400 MHz were observed to generate SBS process. In Ref. [214], a hollow capillary resonator was introduced as a microfluidic device, which could interrogate fluids' optomechanical behaviors through both backward and forward SBSs, and a reasonably sufficient penetration of mechanical energy *could* be achieved from 7 MHz to 11 GHz. In Ref. [215], optomechanical behavior at 60 – 70 MHz from SBS in a liquid droplet was demonstrated and studied. It is worth noting that even through forward SBS could provide optical gain at Stokes frequency, no forward SBS laser has been proposed. Because the corresponding acoustic phonons in aforementioned WGM resonator platforms often have a lifetime longer than that of involved photons, mechanical modes tend more to be coherently excited instead of Stokes optical modes [208,216].

## *3.6 Thermo-optical effect*

Thermo-optical effect refers to the temperature dependence of the refractive index. Such an effect could not be classified as a nonlinear optical effect unless the temperature change is caused by light. However, since the thermal sensor is one important application based on thermos-optical effect, we also review the WGM-resonator-based thermal sensors here. In a WGM resonator, the perturbation of temperature leads to a shift of resonance due to both the thermo-optical effect and the thermal expansion of the resonator structure. The two-fold temperature dependence could be expressed as [217]:

$$\Delta\lambda = \lambda \left(\frac{dn}{ndT} + \frac{dD}{DdT}\right)\Delta T = \lambda(\alpha + \beta)\Delta T \qquad (14)$$

where α and β refer to the thermo-optic coefficient and the thermal expansion coefficient, respectively. Due to the narrow linewidth of high $Q$ WGMs, the detection limit of WGM resonator-based thermal sensor could be smaller than 1 μK [218] and a broad dynamic range from 110 K to 300 K was demonstrated [219]. Apart from silica, fluorite glass, polymer protein material and lithium niobate were also applied to realize sensitive WGM resonator-based thermal sensors [218,220–222]. From the structure perspective, the hybrid coating was introduced to increase the material sensitivity to the temperature signal [223,224]. Polymer encapsulation was also reported to increase the portability of the thermal probe [225–228], and hollow WGM resonators were applied to interrogate the temperature of the liquid sample [229]. In Table 4, the performances of the aforementioned thermal sensors were summarized.

Table 4. Parameters and performance of WGM-based thermal sensors.

| Ref. | Resonator | Q | λ (nm) | \|Sensitivity\| (pm/K) | α (K$^{-1}$) | β (K$^{-1}$) |
|---|---|---|---|---|---|---|
| [217] | Silica Sphere | ~ $10^7$ | 1550 | 15 | $5.5 \times 10^{-7}$ | $8.5 \times 10^{-6}$ |
| [220] | PDMS sphere | ~ $10^6$ | 1480 | 245 | $-7.1 \times 10^{-5}$ | $2.7 \times 10^{-4}$ |
| [223] | PDMS coated toroid | $1.5 \times 10^6$ | 1550 | 151 | $-1.3 \times 10^{-4}$ | $2.7 \times 10^{-4}$ |
| [218] | MgF$_2$ | ~ $10^9$ | 1560 | 0.73 | $7.1(3.0) \times 10^{-7}$ * | $9 \times 10^{-6}$ |
| [222] | LN | $2-3 \times 10^5$ | 1500 | 6.3 | $0.9(-2.6) \times 10^{-5}$ * | $1.2 \times 10^{-5}$ |
| [226] | Polymer packaged silica toroid | ~ $10^7$ | 980 | 131 | / | / |
| [229] | Silica microbubble | $5 \times 10^5$ | 775 | 200 | / | / |
| [221] | Silk toroid | ~$10^5$ | 1434 | 117 | $-1.1 \times 10^{-3}$ | / |

*In crystals, the thermal-optic coefficient could be polarization sensitive.

Heat generated by the optical absorption could also trigger the thermo-optical effect and lead to a broad range of phenomenon. Such a process shares some similarities with optical Kerr effect, but with a much slower response time. The optically induced thermo-optical effect in WGM resonators was first studied by V.S. IIchenko *et al.* in 1992 [230]. Two types of thermal relaxations in resonators were proposed and analyzed, namely, the heat dissipation from modes to the resonator and the heat dissipation from the resonator to the environment, with a typical response time of μs and ms, respectively. In Ref. [231], with the slow scanning of incident laser frequency, optical bistability and hysteresis were observed, as shown in Fig. 11(a). The slow scanning rate assured a heat equilibrium at a temperature deviation ΔT from the original temperature, so that the slow heat dissipation dominated over the fast one. Thus, the resonance wavelength is shifted as in

$$\lambda_r(\Delta T) = \lambda_0 \left(1 + \left(\alpha + \frac{dn}{dT}\right)\Delta T\right) = \lambda_0(1 + a\Delta T) \tag{15}$$

where a is the temperature coefficient considering both thermal expansion and thermo-optical effect. Under such circumstances, the transmission spectrum has the same feature with that in optical Kerr effect, *i.e.*, a triangular shape in transmission during a wavelength up-scanning process, known as thermal broadening. While every high-*Q* WGM resonator encounters a thermal shifting of resonance, the sharp changing in transmission could only be observed with an above-threshold input power [232]. The optically induced thermal nonlinearity enables various all-optical signal processing devices, such as optical switches, diodes *etc.* [233–238]. In addition, relative effects are widely considered and applied during optical experiments of WGM resonators, *e.g.*, thermal locking [231], thermal compensation [239], tuning resonance [240], loss characterization [241].

With a fast scanned incident laser, the thermal equilibrium is replaced by a dynamic described by [16]:

$$\dot{a} + a[\gamma + i(\omega + \delta)] = F(t) \tag{16.1}$$

$$\dot{\delta} + \Gamma\delta = \Gamma\xi|a|^2 \tag{16.2}$$

where a refers to the mode amplitude; $\delta$ is the frequency shift due to the thermo-optical effect; $F(t)$ refers to the incident laser field; $\Gamma$ and $\xi$ characterize thermal dissipation and nonlinearity, respectively. An oscillatory instability could be predicted by solving such equations. Experimental observations of the thermal oscillation were widely reported in WGM resonators [16,230,242–247]. A typical transmission spectrum is shown in Fig. 11(b). Interactions between different nonlinearities, including optical Kerr effect, photorefractive effect, thermo-optical effect (monolithic or hybrid materials) and thermo-optically induced expansion, were also studied [244–248]. When two processes with different time-scale interplay, a slow-fast oscillation form could be observed, as shown in Fig. 11(c).

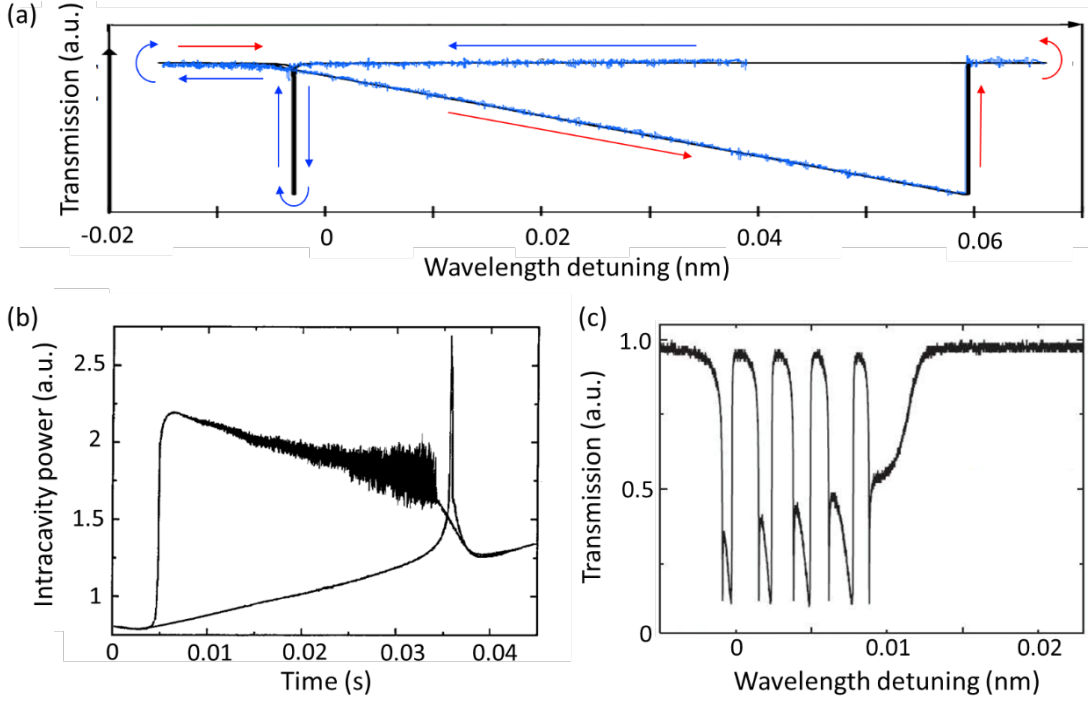

Fig. 11. Different thermal dynamical behaviors in WGM resonators (a) Bistability with slow scanning (heat equilibrium) [231]. (b) Thermal oscillations with fast scanning [242]. (c) Oscillations with multiple thermal-optical nonlinearities [244].

### 3.7 Nonlinearity in Parity-time-symmetric system

In this section, a specific physical platform, namely parity-time (PT) symmetric optical system, will be discussed. As an example of non-Hermitian physics, PT-symmetric systems have been raising considerably attention throughout the fields of optics [68–72,75,249,250], electronics [251,252], atom gases [253–255] and recently were reported as a promising platform to host unique nonlinearity-induced optical effects [68–70,256]. The discussion of PT-symmetry arose from the complexity of the eigenvalues of a non-Hermitian Hamiltonian. While a non-Hermitian system usually has a complex energy spectrum, a PT-symmetric system has a purely real spectrum when it stays in the "unbroken regime" [257]. A physical interpretation of PT-symmetry is that the system remains identical after reversing parity, or the spatial coordinates, and time. It should be noted that for a non-Hermitian system, time reversal means a reversal between loss and gain. Quantum mechanically speaking, the PT-symmetry refers to $[PT, H] = 0$, where the operators $P$ and $T$ are defined as in $P\psi(\mathbf{r}, t) = \psi(-\mathbf{r}, t)$ and $T\psi(\mathbf{r}, t) = \psi^*(\mathbf{r}, -t)$. A typical example of a PT-symmetric system based on WGM resonators is shown in Fig. 12(a) [68,69]. The red (blue) resonator refers to the active (passive) one. With the identical value of loss and gain, denoted as $\Gamma = \Gamma_l = \Gamma_g$, the coupled resonator system satisfies PT-symmetry. The Hamiltonian of the system reads:

$$H = (b_g^\dagger \ b_l^\dagger) \begin{pmatrix} \Omega_0 + i\Gamma_g & \kappa \\ \kappa & \Omega_0 - i\Gamma_l \end{pmatrix} \begin{pmatrix} b_g \\ b_l \end{pmatrix} \qquad (17)$$

where $b_g$ and $b_l$ are annihilation operators for the corresponding modes in resonators with gain and loss, respectively. $\kappa$ is the coupling coefficient between the two resonators. Diagonalization gives the two eigenvalues as:

$$\Omega_\pm = \Omega_0 \pm \sqrt{\kappa^2 - \Gamma^2} \tag{18}$$

When $\Gamma < \kappa$, both eigenvalues are purely real. Although neither resonator is lossless, the strong coupling compensate the loss with gain, realizing two lossless supermodes with a frequency split, known as the unbroken PT-symmetric regime. When $\Gamma > \kappa$, the two supermodes are localized in each resonators, with the same frequency but an opposite-sign gain and loss, known as the PT-broken regime. Between the two regimes, when $\Gamma = \kappa$, the two eigenvalues become degenerate, and the two corresponding eigenstates also merge at such a point, known as the exceptional point (EP). The behavior of both real and imaginary part of eigenvalues are shown in Fig. 12(b) [68]. In Ref. [68,69], nonlinearity induced non-reciprocal transmission was reported in the coupled resonators system and presented lower input power and tunability. Both resonators are fabricated at the edge of the substrates, so that the coupling coefficient $\kappa$ could be easily tuned by varying the coupling gap. Noting that while the ideal real spectrum requires exactly balanced gain and loss, a bias in the imaginary part of eigenvalues will not affect the physical behaviors of the system. In the active resonator, there is a nonlinear gain-saturation. When the PT-symmetry is broken, only the supermode localized in gain resonator has a net gain, inducing a strong field localization regardless of transmission direction. Thus, an asymmetry of optical field distribution is formed, leading to an enhancement of nonlinearity induced non-reciprocity, shown in Fig. 12(c) and will be further discussed in Sec. 5.2. Thanks to the enhancement, the phenomenon shown in Fig. 12(c) only required a probe power as low as 1 μW. In Ref. [256], the field localization in the broken regime was also reported theoretically to a non-reciprocal nonlinear Fano resonance in a similar coupled-resonator system, where two coupled passive resonators with different intrinsic losses presented featured phenomenon of a PT-symmetric system, known as a passive PT-symmetric system. In Ref. [70], a passive coupled dual-resonator was studied, one of which had a tunable loss. Navigating the parameter space via increasing the loss, the injected power into the whole system and the fixed resonator dropped first but revive at a certain point in the unbroken regime and keep increasing deeply into the broken regime. With such non-trivial field behavior, revival phenomena of both thermal broadening and SRS lasing were observed, showing the richness of nonlinear optical behavior in PT-symmetric WGM resonators. While the aforementioned works all showed significant performance in the broken regime, in recent non-Hermitian physics researches, more and more bizarre phenomena were revealed near EP. In Ref. [258], Zhang *et al.* proposed a giant efficient Kerr nonlinearity near EP point. Although it focused on mechanical oscillators, the derivation of nonlinear Hamiltonian is not necessarily limited to mechanics. It provides helpful guiding in further works in nonlinear optics in PT-symmetric platforms.

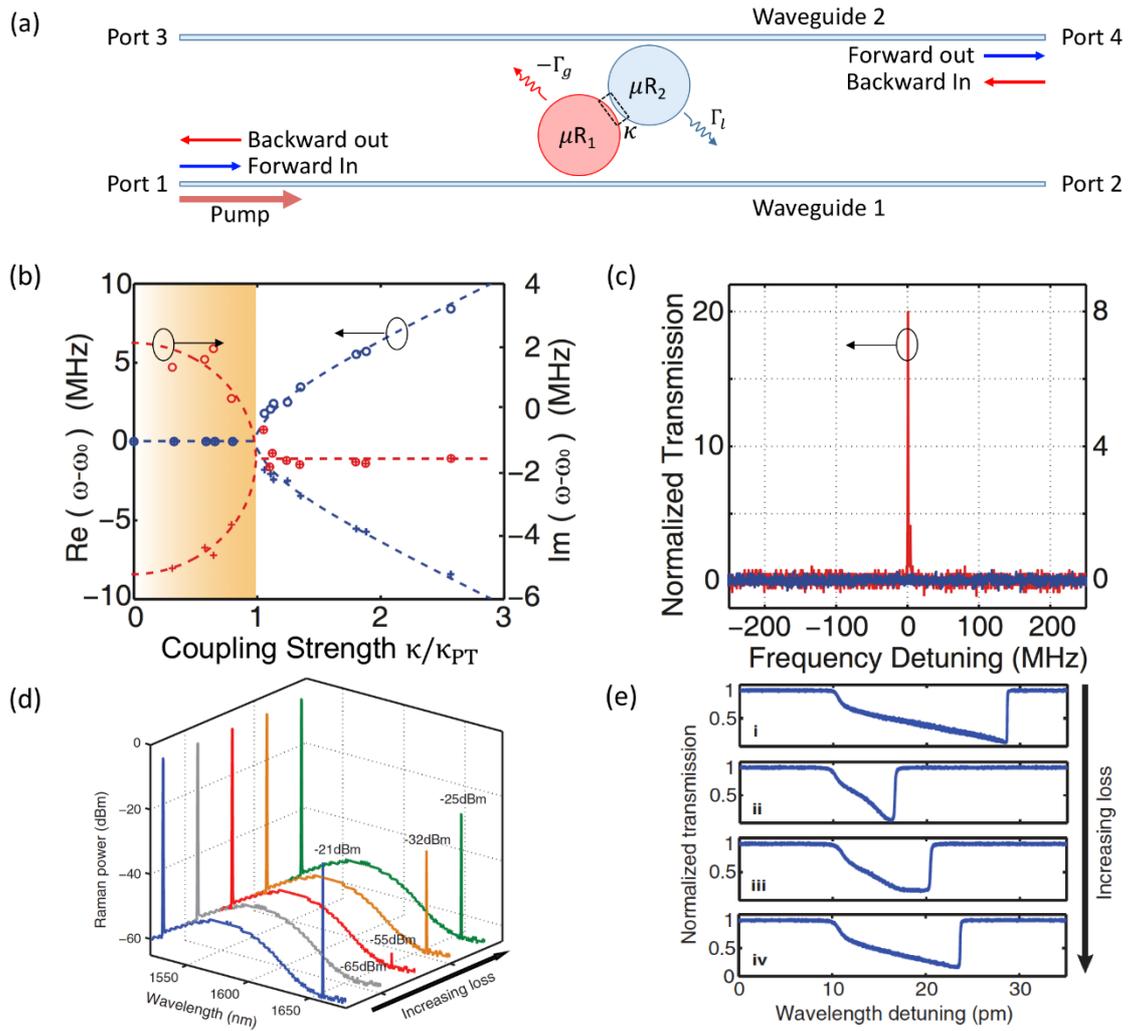

Figure 12. (a) Schematics of parity-time (PT) symmetric coupled resonator system; (b) Real and imaginary part of the resonant frequency of a PT-symmetric system with varying coupling strength [68]; (c) Nonlinear-induce non-reciprocity [68]; (d,e) Revival of Raman lasing and thermal broadening; with increasing loss [70].

## 4. Multi-step nonlinear effects

In this chapter, multistep nonlinear optical effects will be reviewed. With more than one wavelength conversion processes cascaded, a broad range of optical phenomena has been observed and studied. Cascade processes were applied to expand the frequency coverage of nonlinear effects or assist the occurrence of some specific conversions. Here, we will start with one extremely important process, Kerr frequency comb, which is initiated from the cascade FWM. Both the dynamic mechanism and recent progress of Kerr frequency comb generation will be reviewed in detail. Then, hybrid cascade third-order nonlinear processes, mainly containing interactions between SRS, FWM, and THG, will be summarized and discussed.

*4.1 Frequency Comb Generation*

Optical frequency comb (OFC) is a game changer in a broad range of technologies. The 2005 Nobel prize awarded field revolutionized optical clock [259], precise spectroscopy and metrology [260,261], optical waveform and microwave signal generation [262], telecommunication [263] etc. While the first demonstration of optical frequency comb was based on mode-locked laser [264], in the past decade, parametric frequency conversion in WGM resonators with Kerr nonlinearity was applied to generate OFC with a smaller footprint and a larger repetition rate [265]. The microresonator-based optical frequency comb (MFC) originates from the cascade four-wave mixing process. Similarly, as in four-wave mixing, the dispersion-induced non-equidistance of WGMs could be compensated by Kerr effect. However, with normally more than 100 frequency components, the forming of MFC has a much more complicated dynamic. In this section, both theoretical and experimental study on the generation of MFC will be reviewed, starting with a summary of the spectrum spanning of MFCs, followed by a discussion on the noise in the temporal signals, as well as the underlying dynamics, and finally introducing the dissipative Kerr solitons (DKS) and some most recent progress of the different types of phase-locked MFCs.

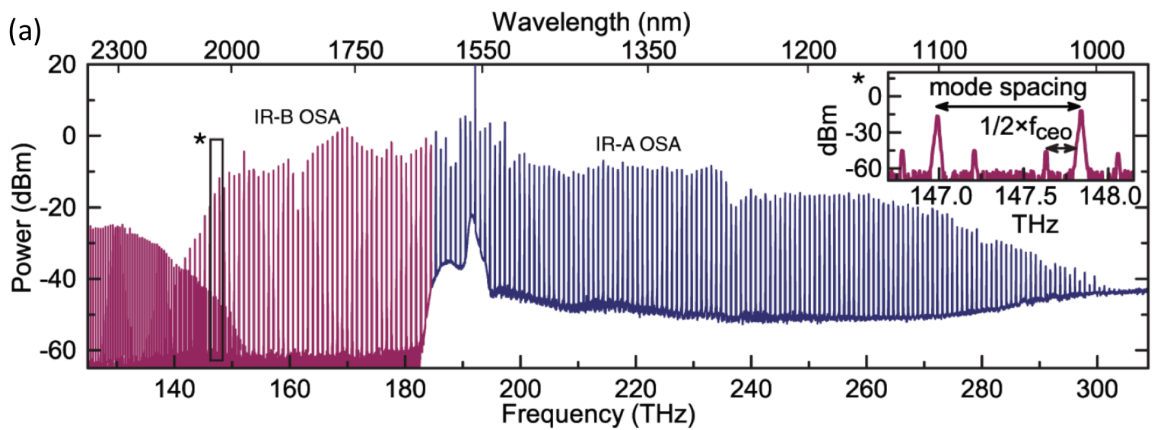

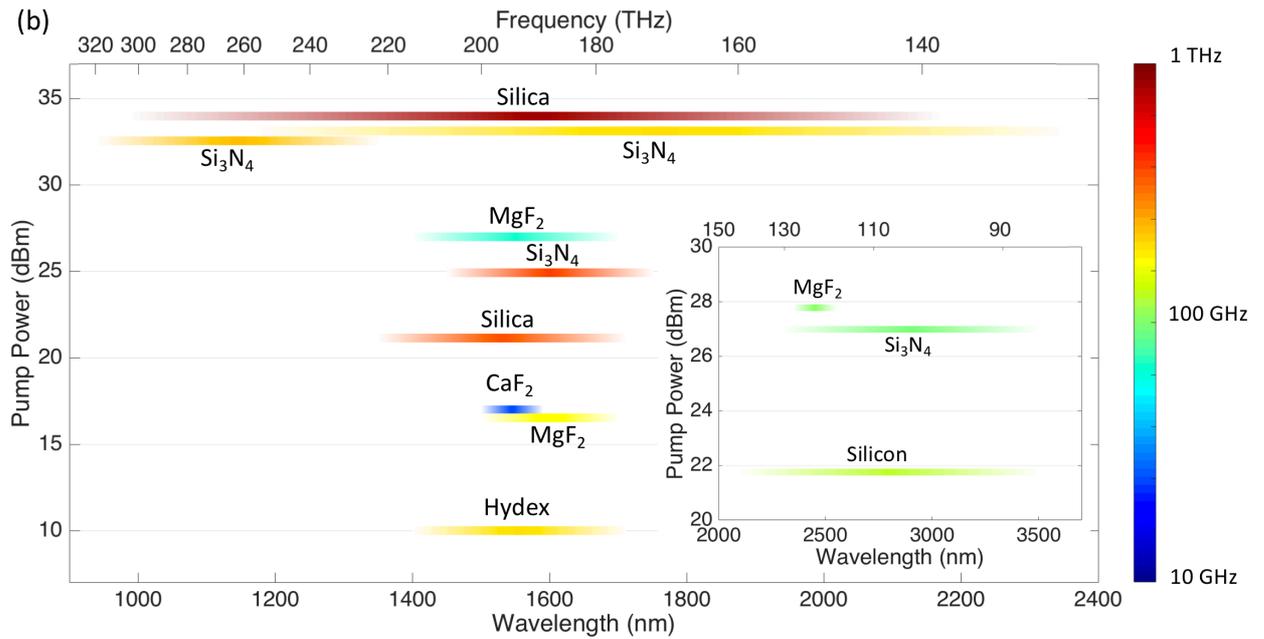

Figure 13. (a) an octave-spanning frequency comb in a silica microtoroid [266], (b) Kerr frequency comb generations in different platforms: pump power, spanning and repetition rate. Reference [174,266–272] (from top to bottom). Inset: Kerr frequency comb generations in mid-IR. Reference: [273–275] (from top to bottom).

Fig. 13(a) shows a typical widely spanning MFC pumped in the telecommunication band. The most critical parameter of a MFC generation is frequency span ($f_{sp}$) and repetition rate ($f_r$), referring to the total spectral range and the frequency spacing between two adjacent teeth, respectively. Additionally, an offset frequency $f_{ceo}$ depicts the deviation of the whole comb from a zero-frequency-start one, which reflects the phase relation between the carrier and the pulse envelope in temporal output. A MFC with a span broad enough such that the maximum covered frequency is twice as the minimum frequency is called octave-spanning. Such a feature allows $f$-$2f$ interference to identify the absolute frequency of the comb. Since 2007, MFCs have been demonstrated in various materials including silica [265,266], hydex glass [177,272], fused quartz [276], fluorite glass ($CaF_2$ [277] and $MgF_2$ [269–271,273], $SrF_2$ [278] crystal) and silicon nitride [174,267,269,279], silicon [275,280] with different WGM resonators including microtoroid [269–271], microbubble [170,281], microring [174,267,269,275,279,280] and crystalline polished structure [269–271,273,277]. In Fig. 13(b), performance and parameter of different works are summarized, while the group velocity dispersion of typical materials could be found in Fig. 6(b). It should be noted that all works shown in Fig. 13(b) are rather early-stage ones focusing on the generation of the comb-shape spectra, which reflect the general features of materials and structures. In silica and hydex, where material dispersion is obviously anomalous near conventional telecommunication band, geometrical optimizations, *e.g.*, minor diameters [265,266], wall thicknesses [170,281], waveguide dimensions [177,272]. In silicon nitride microrings, light fields are strongly confined in the waveguide structure so that the intracavity optical intensity is sufficiently high despite the limited quality factors and provide a strong geometrical dispersion to compensate the normal material dispersion [174,267,279,282]. In crystalline resonators, the rather large structures prevent the significance of the geometrical dispersion. Thus, the comb spanning is largely limited by the material dispersion to no more than 300 nm [269–271,273,277]. However, the flat material dispersion as well as the ultrabroad transparent window make fluorite glasses promising for mid-infrared frequency comb generation [273]. In mid-infrared, silicon has a much smaller nonlinear loss. In Ref. [275], a high-$Q$ silicon microring was fabricated via etchless process and formed a PIN structure for free-carrier sweeping. With 10V reverse bias, a MFC expanding from 2.1 to 3.5 μm was observed.

Another significant challenge is realizing comb in the visible/near-infrared band, where typical materials have strong normal material dispersions. To compensate for unfavorable material dispersions, elaborate geometrical dispersion engineering is required. In silicon nitride microrings, geometrical engineering of waveguide height and width achieved a more than 200 nm spanning MFC pumped at 1064 nm [268]. In silica wedge resonators, with different thicknesses, phase-locked soliton state MFCs were achieved centering at 778 nm, 1064 nm, 1550 nm, respectively [283]. In addition, dispersion engineerings in silica microbubble [170] and in spheroidal crystalline WGM resonator [284] were demonstrated to achieve MFCs with more than 10 lines near 800 nm . While initial four-wave mixing requires anomalous dispersion, MFC generation was also observed in totally normal dispersion wavelength regime [285–294]. Such a counter-intuitive phenomenon could be explained by the revised local dispersion due to mode interaction [286,287,289–294]. Mode interaction could be commonly observed when WGMs from different families approach each other. It was found that the mode interaction, *e.g.*, avoiding of the intersection, could change the local dispersion so drastically that it could reverse the sign of dispersion. Mode coupling could explain a number of non-intuitive features observed, including asymmetric comb shape [286] and the dark pulse formation [289].

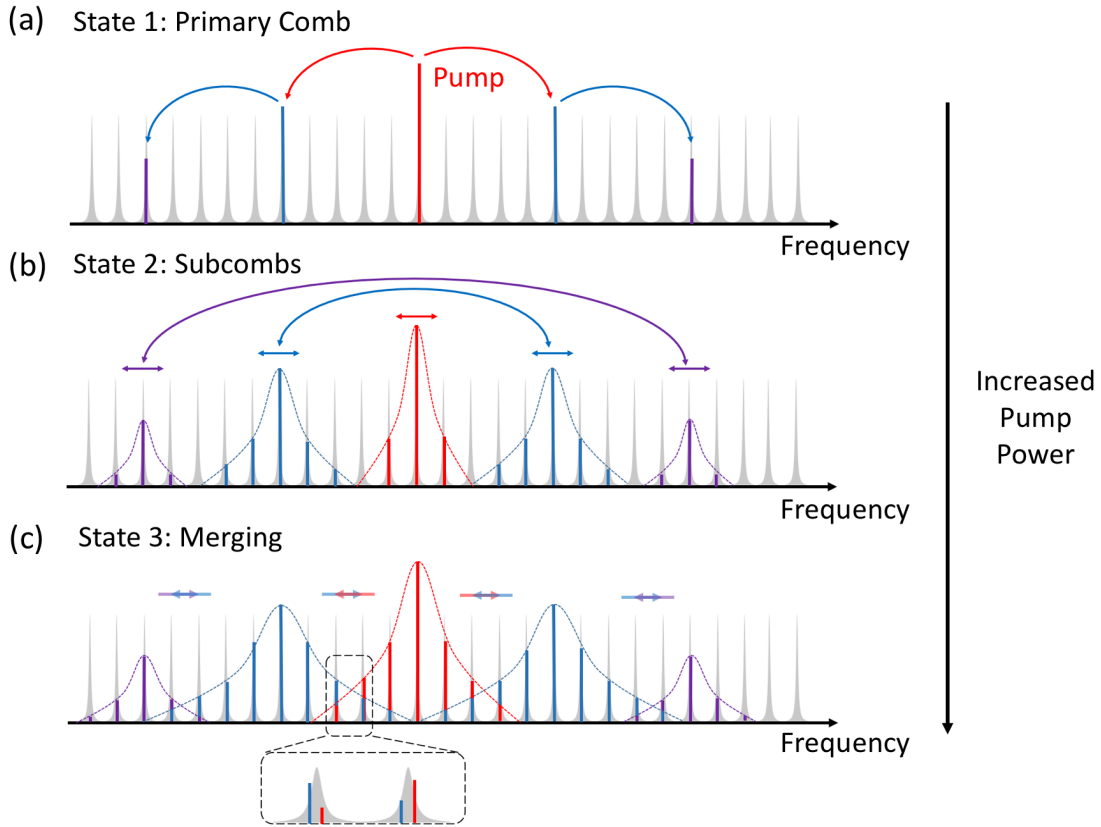

Figure 14. States during the frequency comb generation (multiple mode-spaced, MMS)

However, the magic of frequency combs is not guaranteed by a broadly spanning spectrum. The phase locking among every line is required so that the effective beating between teeth and a sharp pulse train in the temporal signal are available. Unfortunately, most of the early-stage reports of MFC generation, including most of the aforementioned works, were not in the phase-locking state, that is, suffered from the poor coherence of output signal, which had a large RF noise [269,295]. The origin of the noise could be found from the dynamics of the MFC forming. There are two types of dynamics, namely natively mode-spaced (NMS), or type I, and multiple mode-spaced (MMS), or type II. In NMS, comb expands out step by step with a spacing of single FSR, while in MMS, the first generated line pair occurs multiple FSRs away. Lots of broadly spanned MFC follow MMS, which lead to the large RF noise [177,269,270,296]. Roughly three states take place during MMS, with increasing pump power injected. Firstly, cascade FWM lines expand a primary comb with multiple mode spacing (Fig. 14(a), state 1). Secondly, subcombs are formed around the lines of the primary comb (Fig. 14(b), state 2). Meanwhile, in the mid-distant area between the primary subcombs, intermediate subcombs could be generated from pre-existing lines (not shown in the figure). Finally, subcombs merge together to form a consistent frequency comb (Fig. 14(c), state 3). However, at the marginal band between subcombs, a WGM could be populated by multiple lines from different subcombs, which do not necessarily to be phase locked and cause output noise [267,269,297]. In Ref. [269], T. Herr *et al.* gave a simplified expression for the primary frequency spacing in the unit of FSR: $\mu_{th} = \sqrt{\kappa/D_2}$, where $\kappa$ is the total resonator decay rate and $D_2$ is the group velocity dispersion parameter. Thus, the selection of forming dynamics could be achieved by dispersion engineering [271]. However, the poor coherence of MMS combs is not inevitable. With properly tuned pump detuning and power, broad noise in beating signal could undergo a sudden drop, realizing fs-level clean pulse train and confirmed to enter a phase locking regime [267,294,297–303].

The underlying physics of the abrupt phase lock transition is hard to explain in the picture of cascade FWM. Alternatively, the temporal modeling of the pulse formation by solving the Lugiato-Lefever equation (LLE) could better describe the regime transition of MFCs [304]. In a WGM resonator, considering a periodical boundary condition, LLE reads [305]:

$$\frac{\partial \psi}{\partial \tau} = -(1 + i\alpha)\psi + i|\psi|^2\psi - i\frac{\beta}{2}\frac{\partial^2 \psi}{\partial \theta^2} + F \tag{19}$$

where the intracavity field $\psi$ is assumed to have a slow-varying amplitude with respect to $\tau$, whose scale is much larger than the photon lifetime of the resonator; $\theta$ is the azimuthal angle along the propagating direction. $\alpha = -2\delta/\Delta\omega$ is the pump laser detuning normalized by the resonance linewidth; $\beta = -2D_2/\Delta\omega$ is the normalized second order dispersion; and $F = \sqrt{8g_0\Delta\omega_{ext}P/\Delta\omega^3\hbar\omega_p}$ is the dimensionless external pump term. The Kerr nonlinearity is absorbed into $g_0$. Different $\alpha$, $\beta$ and $F$ lead to entirely different temporal behaviors of MFCs, including Turing rolls, chaos, solitons etc. [305–309], and the chaos regime refers to the low-coherent KFC observed in lots of experiments. A most interesting state is the soliton state, or DKS, which will be the main topic for the rest of this section.

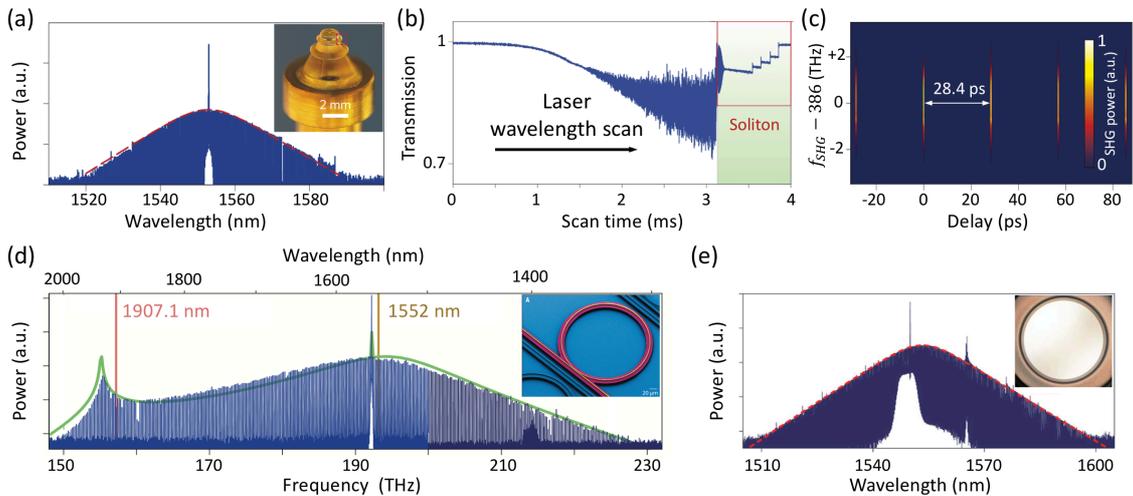

Figure 15. Kerr soliton generations (a)(d)(e) Spectrum of solitons generated in MgF2 [310], silicon nitride [311] and silica [312] WGM resonators, respectively (inset: resonators) (b) Soliton generations is identified by the discrete steps after the pump laser scanning over the resonance mode. (c) frequency-resolved optical gating (FROG) trace of the single soliton state.

DKS is a rather special case of the phase-locked MFCs, where one or multiple isolated pulses circulate the resonator. The intracavity field distribution of N bright DKSs is [310,311]:

$$\psi(\theta) = \psi_{cw} + \psi_1 \sum_{j=1}^{N} \text{sech}\left[\sqrt{\frac{2\delta}{D_2}}(\theta - \theta_j)\right] \exp(i\varphi_0) \tag{20}$$

where the total field amplitude is a superposition of a CW background $\psi_{cw}$ and several sech-shaped isolated pulses with an amplitude $\psi_1$, a phase $\varphi_0$ and a relative angular position $\theta_j$. Experimental demonstration of DKS has been reported in MgF2 resonators [310,313], Si3N4 microrings [311,314], silicon microrings and fused silica disks [315]. In Fig. 15, structures and the output spectra of first works reported in different platforms are shown. Comparing general frequency combs, soliton combs have much smoother envelopes in the frequency domain and high coherence with sharp temporal pulses. The signature of soliton generation in transmission spectrum is discrete steps during pump detuning scanning, shown in Fig. 15 (b). Each step refers to changes in soliton number. In Ref. [316–318], the formation dynamics of DKS was comprehensively studied. DKS generation in the visible band [283], biological imaging window [319], and MIR band [280] were also reported. On the other hand, DKSs occur with a pum at the red-detuning side of the center WGM, which is thermally unstable. To stabilize DKS generation, methods of pump tuning rate control [310], external heating [320], "power kicking" [311,315,321], feed-back-controlled pumping [322], and single sideband modulating [323] were reported. In microrings and disks, where the total dispersion is strongly perturbed from the material dispersion, the dispersive wave could be observed from the output spectra, shown in Fig. 15(d). The dispersive wave originates from zeros of total dispersion, which could be caused by structural dispersion [311,324] or mode interaction [312,325]. The dynamics of the dispersive wave also raises the attention towards third- and higher-order dispersion, which largely determine the shape and spanning of MFCs when GVD is well engineered near zero [177,311]. With broad spanning, high order dispersion was found responsible for the asymmetric spectral and temporal envelope as well as the spectral recoil of center frequency and the dispersive wave peak [177,311,326,327]. In addition, Raman effects was observed in DKS generation. With stronger intracavity circulating power, intrapulse Raman scattering (IRS) could induce a red-shift and a broadening of the soliton spectrum [328,329]. The generation DKSs with a deterministic soliton number was realized by a follow-up back scanning of the pump frequency [330], or by leveraging the spatial mode-interaction [331]. In Ref. [332,333], counter-propagating solitons was reported, whose phase and repetition time could be independently controlled. In Ref. [334], multiple solitons with a crystal-like behavior was observed and studied.

Apart from the generation of stable bright Kerr soliton, several recent experimental works on special Kerr solitons in WGM resonators are noteworthy. In Ref. [289], dark solitons, with temporal dip shapes, were reported in a Si3N4 microring within a band with anomalous dispersion. Breather solitons, which refer to time-varying Kerr solitons, are an important class of unstable soliton states predicted by LLE. Breather soliton was theoretically found related to the Fermi-Pasta-Ulam (FPU) recurrence, an analogue to Kuznetsov-Ma and Akhmediev breathers [335,336]. The experimental observations of breather soliton in WGM resonators were reported very recently, in silicon nitride microrings [337–339], MgF2 crystalline resonators [335,339], and silicon microrings [337]. The signature of breathing solitons could be confirmed from a peak in RF signal and periodically varying soliton power in temporal output. While the maps of multiple regimes in parametric space vary from platform to platform, breather regime usually lies adjacent to stable soliton regime. Thus, studying of breather soliton is significant for understanding and ensuring stable generation of stable Kerr solitons.

*4.2 Other cascade wavelength conversion*

In a high-$Q$ WGM resonator, SRS, FWM, and THG could occur simultaneously [340], competitively [166,341,342], and also cascadedly. In this section, cascade processes among SRS, FWM, and THG are discussed, which are schematically shown in Fig. 16.

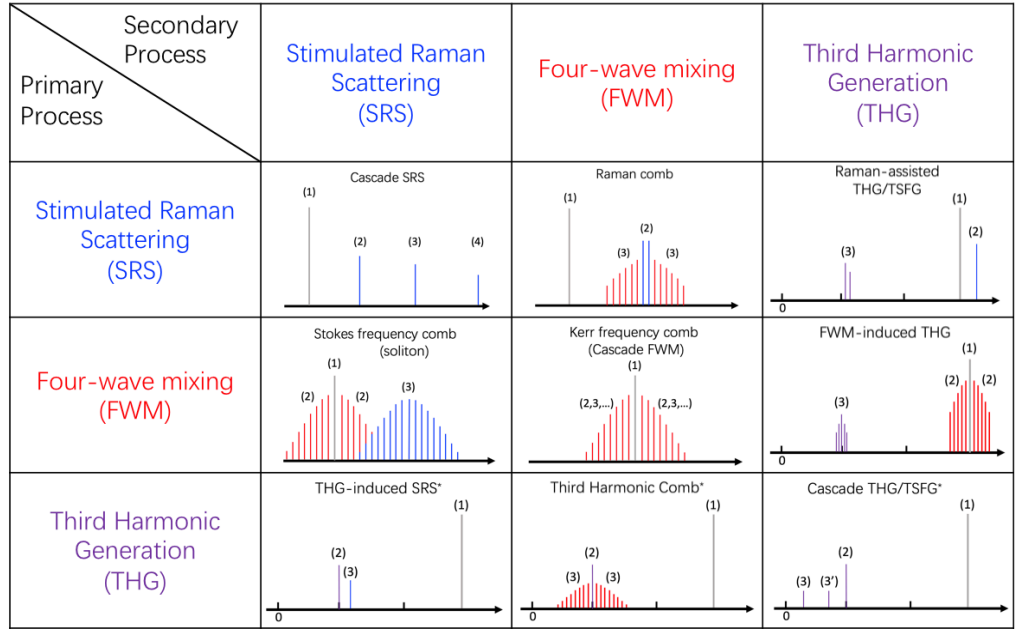

Figure 16. Schematic of multistep nonlinear effect. Numbers in each cell refer to order of generation. (*) not yet experimentally observed.

Since SRS enjoys an intrinsic phase-matching, it commonly serves as a primary process to induce or assist a secondary process. Raman assisted FWM was observed in early-stage SRS researches with liquid droplets [185], where mutual FWM occurs among original SRS lines and expand the coverage of Stokes lines. Similar phenomenon was also reported in solid WGM resonators including silica microspheres and microtoroids [181,186]. It should be noted that two Stokes photons with different frequencies could also interact with one pump photon to generate sidebands around pump lines despite the unfavorable local dispersion [343]. A typical experimentally observed spectrum of Raman-assisted FWM is shown in Fig. 17(a) [343]. Raman Stokes lines form a comb spectrum, namely Raman comb, and sidebands are also generated around the pump. Theoretical analysis of the Raman comb generation is available in the time domain [344–346], and experimental observation of low-noise phase-locked Raman comb was reported in fluorite glass WGM resonators [343,345]. Raman-assisted THG/TSFG was widely observed in both liquid droplet resonators [154] and solid WGM resonators [153,158,347,348]. Multi-color output could be realized in this way, shown in Fig. 17(b). With further increased pump power, cascade SRS and Raman comb could also participate into

THG/TSFG process, achieving an even broader emission spectrum.

With a well-engineered dispersion, FWM could have a lower threshold than SRS [166], allowing it to serve as a primary process. In mid-infrared frequency comb generated in silicon microring resonators, SRS was found to assist the mode- locking transition to achieve low-noise coherent state [302]. When SRS happens between different mode families, an initial soliton frequency comb could generate another Raman-shifted soliton, known as Stokes soliton [349], whose spectrum is shown in Fig. 17(c). FWM in the infrared band could also induce THG/TSFG for visible emissions [348,350,351]. FWM-induced process requires finer dispersion engineering but could achieve a broader emission band, even forming a visible comb [350] (Fig. 17(d)).

Third harmonic generation is weaker than SRS and FWM, which makes no cascade process whose primary one is THG has been observed. Yet we still list the possible processes in Fig. 16, some of which could be promising in specific applications. For example, cascade THG/TSFG could be applied to generate ultraviolet emission.

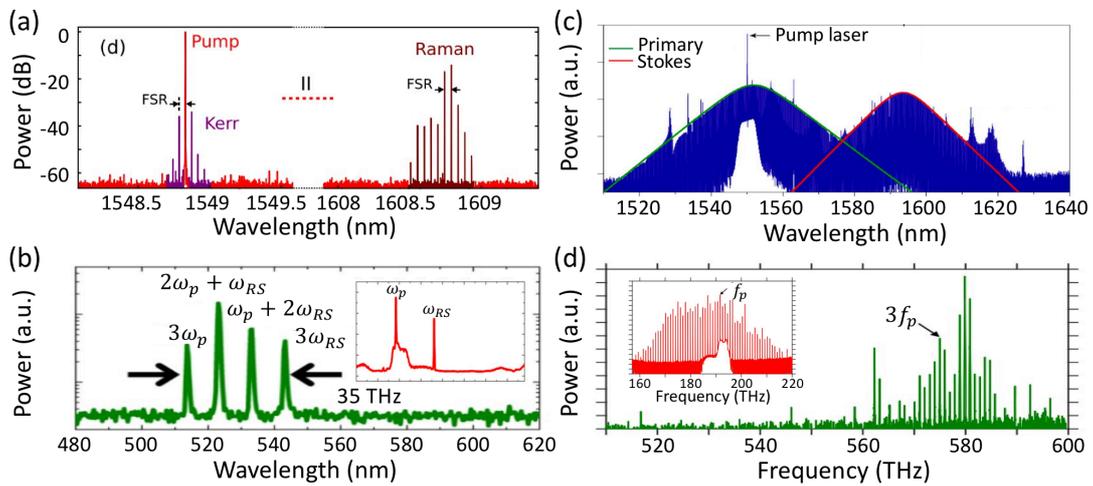

Figure 17. Experimental spectra of (a) Raman comb [343]; (b) Raman-assisted third harmonic generation/third-order frequency sum generation (THG/TFSG) [348]; (c) Stokes Kerr soliton [349]; (d) FWM-assisted THG/ TFSG [350].

Apart from cascade processes among SRS, FWM, and THG, other types of cascade nonlinear process have been reported. In Ref. [161], SHG and second-order SFG was cascaded to generate third-harmonic emission ($\omega_{TH} = \omega_{SH} + \omega_P$). Cascade SBS and FWM was also observed in fluorite glass WGM resonators [352] and silica microbottle resonators [209]. In Ref. [353,354], a MFC centered at telecom band stimulate combs near the visible band (~765 nm) through second order nonlinearity.

## 5. Applications

### 5.1. Narrow-linewidth laser

A low-noise laser is a basis to a broad range of applications, including metrology, spectroscopy and fundamental science. Narrow-linewidth lasers have been reported in WGM microresonators made from Erbium-doped silica (population inversion) [355] and pure silica (SRS and SBS) [13,356]. Moreover, highly coherent microwave signals could be generated by heterodyne of two low-noise laser lines. Such an optical scheme of microwave synthesizing shows promise in overcoming the scaling problems in electronic microwave generators [357]. Micro-lasers based on the backward SBS in WGM resonators is especially fit for achieving narrow-linewidth lasers and the GHz level Brillouin shift allows cascade SBS to be an ideal source of microwave synthesizing. In this section, both general theory and experimental progress of SBS low-noise laser and microwave synthesizers in WGM resonators will be reviewed.

In a SBS micro-laser, two fundamental noise sources are considered, namely pump noise and Schawlow-Towns (ST) noise [207,357,358]:

$$\Delta\nu = \frac{\Delta\nu_p}{(1+\Gamma_A/Y_B)^2} + \frac{\hbar\omega^3}{4\pi P_{out} Q_{tot} Q_{ext}}(n_T + N_T + 1) \quad (21)$$

In the first term, $\Delta\nu_p$ is the linewidth of the pump laser, and $\Gamma_A$ and $Y_B$ are the damping rates of the acoustic mode and the optical Stokes mode respectively. $\Gamma_A$ is usually much larger than $Y_B$, leading to a strong suppression of pump noise. Thus, an SBS laser could significantly clean up the noise in pump laser, achieving quantum-limited-noise level lasers at frequencies of user's option. The second term in Eqn. 21 refers to the quantum noise, or ST-like noise for it is inversely proportional to the output power $P_{out}$. $Q_{tot}$ and $Q_{ext}$ are the total and the external $Q$ factors. $n_T$ and $N_T$ refer to the numbers of thermal quanta in the mechanical field and the optical field. In a WGM resonator, $Q$ factors could exceed 108, reducing the ST linewidth down to sub-Hertz level.

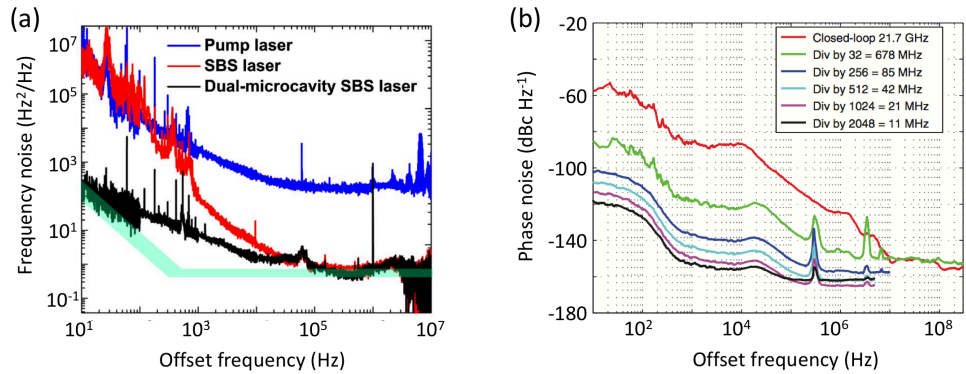

Figure 18. (a) Frequency noise of the pump laser, SBS laser, and dual-microresonator SBS laser [359]. (b) Single-sideband phase noise (L(f)) of microwave generation from cascade SBS laser in a microresonator [357].

The linewidth of an SBS microlaser in WGM resonators was first measured in Ref. [13]. A 0.06 Hz2/Hz of frequency noise was observed and was verified to be dominated by the ST noise [360]. In Ref. [361],low-noise SBS laser at 1064 nm was demonstrated in the same platform. ST frequency noise of 0.1 Hz2/Hz at 1.8 mW output power [361]. With a narrow-linewidth SBS laser, more accurate gyroscope could be demonstrated [362]. According to Eqn. 21, ST noise in SBS laser is dominated by the number of mechanical thermal quanta $n_T$. The relation between the linewidth of SBS microlasers and the temperature was verified from room temperature to cryogenic temperature [363]. Furthermore, noise at low frequency, which is limited by the thermos-refractive noise of the resonators could be suppressed by a dual-cavity configuration [359], (Fig. 18(a)). On the other hand, highly coherent microwave synthesizer could be developed based on photo-mixing of cascade SBS lines. In Ref. [357], low-noise K-band signal at 21.7 GHz was generated out of 1st and 3rd SBS Stokes lines (Fig. 18(b)). As the noise of SBS lasers is insensitive to the corresponding optical frequencies and orders of SBS, the phase noise of generated microwave signal will not degrade with increasing frequency, which largely overcomes the scaling problem of its electrical counterpart.

*5.2 Nonlinear induced non-reciprocal and chiral devices*

Non-reciprocal optical devices could achieve unidirectional transmission, which is critical for optical isolators for laser protection, and optical diodes/circulators in signal processing [364]. However, the asymmetric transmission between two ports of interest is not equivalent to the non-reciprocity. While the general case of non-reciprocity was discussed in Ref. [365], here we could safely simplify the problem by requiring all ports in our system to be single-mode. In this way, the total scattering matrix could be reduced to 2-by-2, and the non-reciprocity is equivalent to asymmetric transmission between the two ports. In Fig. 19(A), the reduced model only containing the ports of interest is presented. An ideal non-reciprocal device refers to a scattering matrix as asymmetric as:

$$\begin{pmatrix} b_1 \\ b_2 \end{pmatrix} = \begin{bmatrix} 0 & 0 \\ 1 & 0 \end{bmatrix} \begin{pmatrix} a_1 \\ a_2 \end{pmatrix} \tag{22}$$

Where $a_{1,2}$ and $b_{1,2}$ are the input and output field amplitude at the two ports, respectively. It could be proved that the symmetry of the scattering matrix could be inferred by Lorentz reciprocity theorem $\nabla \cdot (E' \times H'' - E'' \times H') = 0$ ($[E', H']$ and $[E'', H'']$ are Electromagnetic fields of two arbitrary group of excitation). The Lorentz reciprocity is built on the assumption of symmetric, time-invariant and field-independent permittivity **ε** and permeability tensor **μ** [365]. Thus, to break the reciprocity, one must attack at least one of those properties. Magneto-optical materials could be employed to realize asymmetric permittivity with external magnetic field [366], and time-dependent refractive indices were reported to achieve optical non-reciprocity [367]. In addition, optical nonlinearity could play a crucial role in non-reciprocity, for it could provide field-dependency to the permittivity. More specifically, the key question is how to realize different permittivity for opposite transmission directions. Most of the relative works could be classified into two types, namely structural selectivity and phase-matching selectivity.

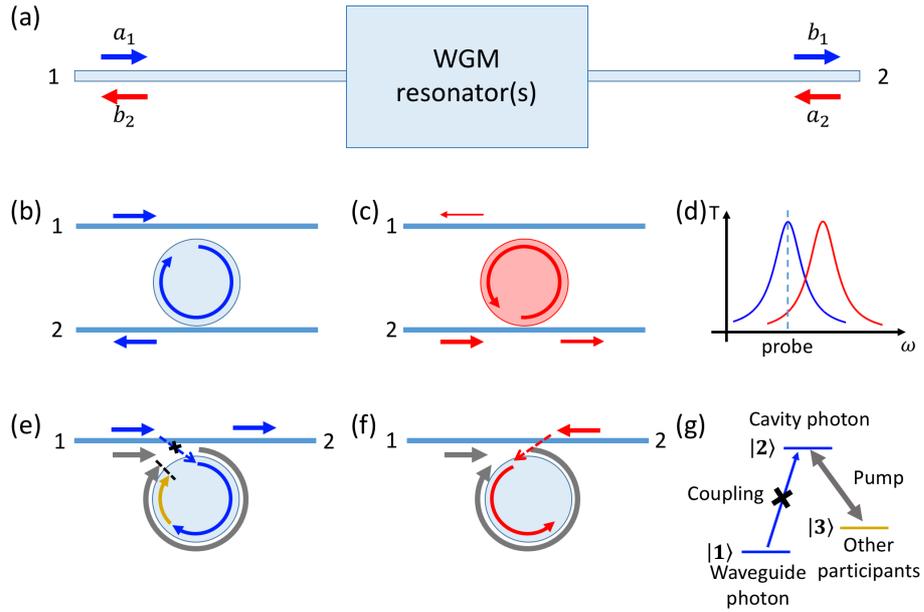

Figure 19. Schematics of nonlinear induced non-reciprocity. (a) Forward transmission (blue) is from port 1 to port 2; Backward transmission (red) is from port 2 to port 1. Brown region abstracts the interaction with resonators. (b)(c) Bidirectional transmission process under structural selectivity. Blue and red resonator refer to different intra-cavity light intensity. (d) Transmission spectra of the forward and the backward case under structural selectivity. (e)(f) Bidirectional transmission process under structural selectivity. Gray arrow refers to pump light. (g) State diagram of an optical analogue of Electromagnetically induced transparency (EIT).

Structural selectivity: As shown in Fig. 19(b-d), when asymmetric structures could lead to different intracavity energy with the same input from different directions. Nonlinear effects will be triggered variously, causing different index shift due to Kerr-type nonlinearity. Therefore, non-reciprocal transmission could be achieved. In Ref. [367], Fan *et al.* reported a passive optical diode in silicon microring platform, whose non-reciprocity was initiated by the thermo-optical nonlinearity in an asymmetric add-drop filter (ADF). The asymmetric transmission was further amplified by the notch filter (NF), which could resonantly absorb the attenuated light in backward direction but could not block the forward light due to the thermal red-shifting of the resonance. With an optimization of fabrication, such a scheme could achieve a non-reciprocal transmission ratio (NTR) of 40 dB with ~2.3 mW incident power [238]. Asymmetric structures could also realize direction-dependent nonlinear optical, provided by the gain saturation in active WGM resonators [68,69,368]. In Ref. [368], the asymmetric coupling changed the saturation intracavity field amplitudes and subsequently gain coefficients of different directions. By tuning the pump power and coupling coefficients, non-reciprocal transmission could be observed with an input power ranging from 10 nW to 100 mW, and the insertion loss of ADF could be compensated by the gain. The performance of such a scheme could be further improved even combined with the PT-symmetric platform, shown in [68,69]. The strong localization in the broken-phase coupled resonators could provide an enhanced and flexible asymmetric optical structure to host the nonlinear effects.

However, for structural selectivity, intracavity optical energy is the critical factor to distinguish different transmission directions. Thus, they could only operate with an input signal from a single direction. What's more, such non-reciprocal performance could not fit the requirement for an optical isolator, where both a strong forward signal and a weak backward scattering exist [369]. Going beyond such limitation, another type of nonlinear non-reciprocal should be mentioned, which is based on phase-matching selectivity.

Phase-matching selectivity: As shown in Figs. 19(e) and (f), within a WGM resonator, given a strong pump light circulating in a specific direction, probe lights traveling in opposite directions could witness different phase-matching conditions, thus initiating non-reciprocal nonlinear effects. To realize non-reciprocal transmission, the phenomenon of induced transparency is usually applied [370–374]. Shown in Fig. 19(g), the induced transparency in a nonlinear resonator could be described by a lambda configuration analogue. The inward coupled photon in the anti-stokes resonance refers to the absorption transition between states |1⟩ and |2⟩. State |3⟩ represent another participant quanta involved in the nonlinear wavelength conversion. When the phase is matched with input from one direction, the fast oscillation between |2⟩ and |3⟩ will prevent the transition between |1⟩ and |2⟩, inducing transparency of the on-resonance WGM, while efficient coupling or "absorption" into the WGM is maintained with input from another direction. Classified by the type of state |3⟩, Brillouin induced transparency (BSIT) and parametric nonlinear optically induced transparency (NOIT) could be realized. In BSIT, the phonon state of the traveling acoustic mode serves as |3⟩, which is similar to that of OMIT. A strong pump light or control light is set to be at the lower frequency, and the induced transparency is probed at anti-stokes frequency [370,371]. BSIT requires that the lifetime of phonons should be much longer that resonance photons [370]. Thus, only forward SBS could trigger BSIT. In Ref. [372], NOIT was achieved based on second-order parametric wavelength conversion in an AlN mirroring resonator. During the process, two telecom frequencies ($\omega_a$ and $\omega_c$) and one visible frequency ($\omega_b$) were involved, satisfying $\omega_a + \omega_c = \omega_b$. The Pump at $\omega_c$ trigger the oscillation between photon at $\omega_a$ and $\omega_b$ (|3⟩) was formed. Then, the original Lorentz dip at $\omega_a$ was altered into induced transparency lineshape. In addition, non-reciprocity could also be generated by the phase-matched four-wave mixing. Pumped below the lasing threshold, EIT-like [373,375–377] and Fano lineshape [374,378–380] in the transmission spectrum were observed non-reciprocally.

Apart from transmission problems, reversal symmetry could be broken in terms of chirality generation, or the breaking of intrinsic mirror symmetry. In WGM resonators, mirror-symmetry exists between clockwise (CW) and counterclockwise (CCW) modes, also reflected by the identical ratio of lasing behavior in both directions. To generate chirality, normally the mirror symmetry of the resonators need to be broken through deformation, scatterer, rotation [381] or more recently, exceptional point tuning [381] *etc*. In 2017, two papers were published, in which spontaneous chirality was achieved without spatially breaking the mirror symmetry [382,383]. In both papers, the chirality was initiated by Kerr-nonlinearity. The balanced counter-propagating light pair coupled through CPM. When the intracavity intensity exceeded a threshold, the balance became unstable and transferred into a chiral mode [382], or in other words, a small deviation from the balanced CW and CCW modes underwent a self-amplification [383]. Recently, by altering the optical setup, such chirality could be translated into non-reciprocity and realized optical isolators and circulators [384].

## 5.2 Applications of MFC

In Sec. 4.1, the progress of MFC and DKS generation has been discussed. However, to better understand the motivation and the optimization goals of those works, it is still necessary to review the main applications of frequency combs in WGM resonators. It should be noted that most of applications require fully phase-locked comb state, which could be reliably achieved through DKS pathway. Thus, most of works mentioned in this section were based on Kerr soliton generation or MFCs in the phase-locked state.

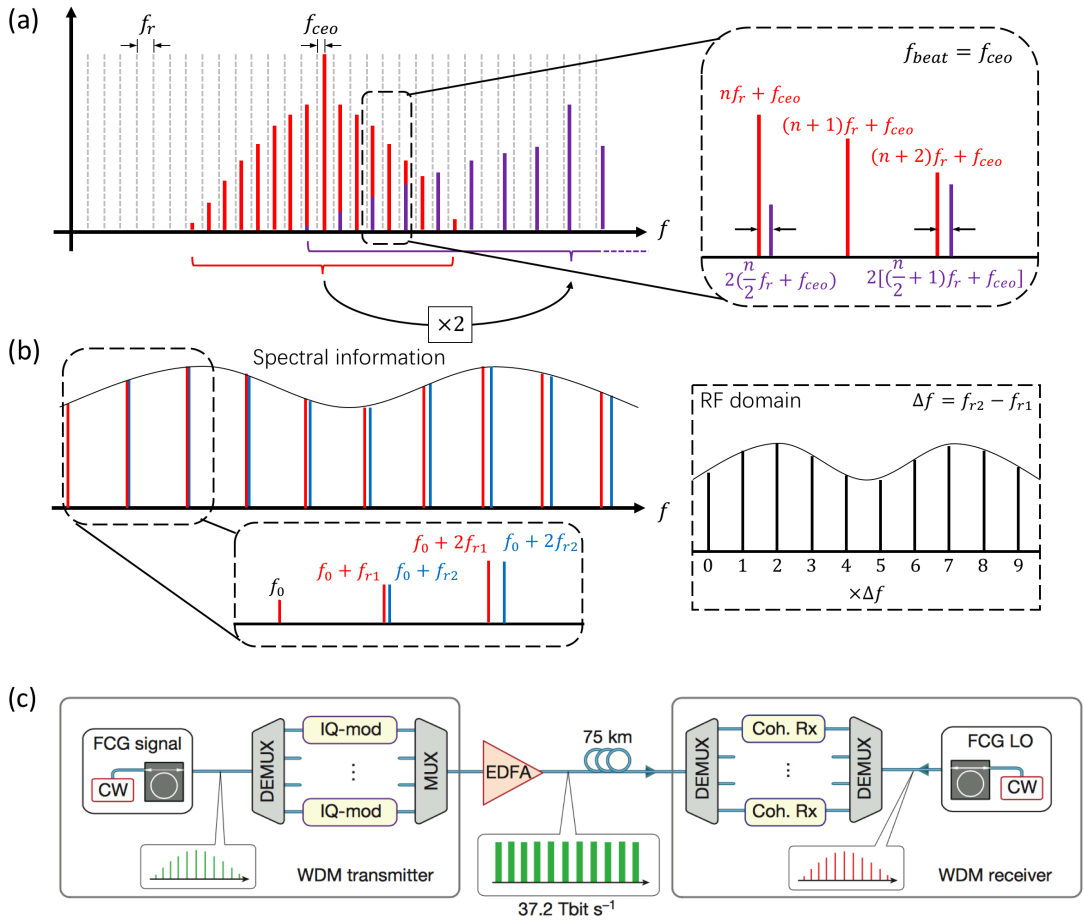

Figure 20. (a) Schematics of the f-2f interferometer and the link between optical frequency and radio frequency. (b) Schematics of dual-comb configuration. (c) Scheme of coherent data transmission using WGM resonator based Kerr solitons at both the transmitter and the receiver.

Frequency metrology is the field revolutionized by the advent of optical frequency combs, for OFCs could provide a broad frequency standard with an unprecedented precision [385]. Recalled from Sec. 4.1, the frequency of the comb line with an index of $n$ reads

$$f_n = f_{ceo} + nf_r \quad (23)$$

To achieve all the advantage of OFC-based frequency metrology, both $f_{ceo}$ and $f_r$ in MFCs need to be measured and stabilized. $f_r$ could be interrogated by sending the comb output into photodiode for spectrum analyze and $f_{ceo}$ could be measured with external frequency reference [386–389] or through self-reference process [390–392], both via heterodyne interference. In a mf-kf self-reference (m,k integers), the comb is sent through nonlinear crystals generating $m_{th}$ harmonic and $k_{th}$ harmonic respectively. Then $f_{ceo}$ could be retrieved from the output RF/microwave spectrum which contains the beating of harmonic lines. A schematics of *f*-2*f* self-reference is shown in Fig. 20(a). In Refs. [390–392], self-reference of MFC [390,391] and DSK [392] was reported. As a versatile frequency standard, OFC was proved to be promising in metrology ranging from calibration of astronomical spectrographs [393,394], absolute measurement of atomic lines and laser frequencies [264,395] to the measurement of resonator dispersion [396]. On the other hand, OFC provides an absolute link between microwave and optical frequency, thus allowing the reading out or synthesizing of ultra-high frequency [259,397], which is beyond the reach of electronic devices. In WGM resonators, the demonstration of the MFC-based optical clock was reported in Ref. [388], where a low-noise MFC in silica microdisk divided the 3.5 THz Rb transition $10^8$-fold into 33 GHz repetitive clock output with $5\times10^{-11}/\sqrt{\tau}$ stability. An optical-frequency synthesizer in an integrated photonics platform has also been reported. With two interlocked Kerr solitons with GHz-level (in a silica microdisk) and THz-level (in a $Si_3N_4$ microring) $f_r$, a 10 MHz clock signal was multiplied thorough a frequency chain by more than $1.9\times10^7$ and guided a III/V Si tunable laser in C-band to operate at absolute optical frequencies, with the synthesis error constrained to $7.7\times10^{-15}$ [398]. In Ref. [399], DSK was employed for calibrating the astronomical spectrographs as the first in-the-field demonstration of microcombs.

Another significant killing application of OFC is the precise spectroscopy. The densely and equidistantly distributed frequency lines could serve as an ideal sampling function for one-time interrogation of sample's absorption spectrum. To read out the spectral information, both direct optical spectrum analysis via dispersive spectrometer [260,273] [400], FTIR [401–403], and heterodyne down-mixing to RF spectrum via dual-comb [404–406] are available. Dual-comb spectroscopy provides a spectrum interrogating scheme which is free of tunable lasers and spectrometers and has been realized with MFCs [405]. With two mutually locked soliton combs as reference and signal whose repetition frequencies $f_{r1}$ and $f_{r2}$ are slightly different, the interferogram between the dual-comb one-to-one maps the optical spectral information down to the RF band with a spacing of $\Delta f_r = f_{r1} - f_{r2}$ (Fig. 20(b)). In Ref. [405], Dual-DKS were generated and used to measure the absorption spectrum of the $H13CN\ 2\nu_3$ band. Dual-comb spectroscopy or source was also demonstrated in silicon [400,406], $MgF_2$ [407] and silicon nitride [408] WGM resonators. The idea of dual-comb metrology could be also employed in the ranging system (LIDAR) [409]. The interferogram could translate the fast pulse sequence down to electrically readable signal to infer the target distance. Meanwhile, the ambiguity range of dual-comb LIDAR could be enhanced beyond the pulse-to-pulse distance of the single comb pulse train [409–411].

Apart from applications in metrology, OFC could also serve as an ideal light source for microwave photonics and communication. Once stabilized to an ultrastable CW laser oscillator, the low phase noise of the reference could be transferred to every frequency lines [412]. In WGM resonators, low-phase-noise microwave to millimeter wave, or even longer wave signal generation with an optical-frequency carrier was reported in both low-noise MFCs [297,298,313] and soliton combs [310,311,315,413–415]. Line-by-line pulse shaping was reported in MFC with a liquid-crystal modulator array [289,295]. In communication, OFCs provide single laser pumped, highly coherent, equidistant and narrowband optical carriers for wavelength-division multiplexing (WDM). Especially in minimized WGM resonators, such ideal source could be realized with compatibility and user-defined channel width more than tens of GHz. Terahertz data rate communication was reported in MFC in both low-noise states [313] and soliton state [263]. In the latest work, a 55.0 Tbit/s line rate was achieved. At the same time, further optimization in MFC source was also reported to pave the way for more practical MFC-based optical communication [416–421].

## 6. Conclusion and perspectives

This review article has summarized all kinds of nonlinear optical effects in WGM resonators as well as their applications. These resonant nonlinear effects can be classified into two groups: controlling and generation.

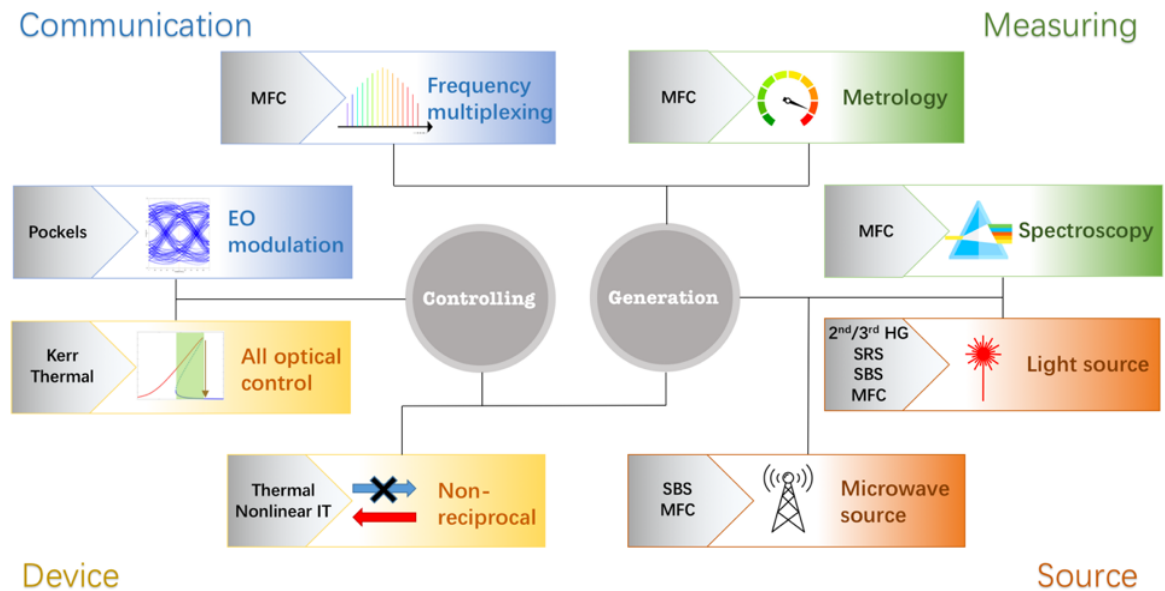

Figure 21. Summary of nonlinear optics in WGM resonators. (Nonlinear IT: nonlinear induced transparency; HG: harmonic generation)

The functions and applications enabled by the two groups as well as the involved nonlinear effects are summarized in Fig. 21. Nonlinear optical controlling could originate from the phase modulation provided by Pockels effect, Kerr effect and thermo-optical effect. On the other hand, a great number of nonlinear optical generation, mostly based on wavelength conversion, could be realized in WGM resonators. Compared to conventional light sources, optical generation from nonlinear optical effect could provide advantages like low noise, freely-selected wavelength, and broadband phase locked comb spectrum. Also, the nonlinear optical generation could also be applied to realize non-reciprocal devices via nonlinear induced transparency; microwave generation via beating; and the link between optical frequencies and radio frequencies.

The advantages of using microresonators to study the nonlinear optical effects rise from their strong field confinement and long photon lifetime. Indicated by a large $Q/V$, such a two-fold advantage not only lows the threshold of various nonlinear effects but also provides an opportunity to observe weak light nonlinear effects down to single/few-photon level [422,423]. Limited by the length, in this review, we have only focused on second- and third-order optical nonlinearities possessing much larger nonlinear coefficient compared with the high-order nonlinear effects. Besides, this review doesn't cover the phenomena related to quantum optics since they are contained by other recent review papers [51].

The study of nonlinear optics in WGM microresonator has achieved great progress in the past, and the several directions deserve to be paid more attention in the future.

In terms of materials, novel monolithic material systems have emerged and shown unique nonlinear optical features, such as diamond [179,198], integrated thin-film lithium niobate [86,88–90,106,121,125,128,424], protein [221,425,426], silicon oxynitride [427] *etc*. Also, hybrid material systems involving special highly nonlinear organic materials [102,428–430] also shows great promise. While the compatibility provided by the host materials still holds, the special nonlinear materials could enhance the performance of nonlinear optical effect.

In terms of the structure, apart from the optimization or new design to enhance the interaction of light and nonlinear materials, the driving force of the future nonlinear optical research could originate from the special structures that could improve phase matching/dispersion engineering and broadband coupling. Phase matching/dispersion engineering is critical to enable a broad range of nonlinear optical effects. Structural design in morphology [169,170,431–433] and coating [223,239] have been triggering a series of nonlinear optical demonstrations. What's more, recent progress in integrated nonlinear photonics, including metamaterials, could be found translatable to WGM resonator platforms for phase matching engineering [434] or even phase-matching-free nonlinear transition [435]. On the other hand, due to the broadband nature of lots of nonlinear optical effects, a versatile coupling channel covering the whole frequency range is highly desired yet challenging. Recently, chaotic modes in weakly deformed WGM resonators were found able to realize an ultra-broadband efficient coupling between a single bus waveguide and WGMs, which could be anticipated to trigger a series of nonlinear optical devices with high output yield [53].

Finally, from the practical perspective, the translation from benchtop nonlinear optical demonstration to the practical device should be paid more and more attention in the future research. Varied with specific applications, case-sized, board-sized, chip-sized devices are all desired. Some noteworthy subjects include the minimization and the robustness of the nonlinear devices and the optical coupling region, as well as the simplification of the whole driving, controlling and processing modules. Such improvements could maximize the advantage of systems based on nonlinear WGM resonators in collaborations between it and fundamental researches as well as human daily practice.

unidirectional laser emission," Laser Photon. Rev. **10**, 40–61 (2016).